\documentclass{aa}  
\usepackage{graphicx}
\usepackage{txfonts}
\usepackage{hyperref}
\hypersetup{colorlinks=true, linkcolor=cyan, citecolor=blue, urlcolor=cyan}
\usepackage{xcolor}

\begin{document}

   \title{The history of star-forming regions in the tails of 6 GASP jellyfish galaxies observed with the Hubble Space Telescope}


   \titlerunning{Star-forming regions in the tails of jellyfish galaxies}

   \author{A. Werle\inst{1}
          \and
          E. Giunchi\inst{1,2}
          \and
          B. Poggianti\inst{1}
          \and
          M. Gullieuszik\inst{1}
          \and
          A. Moretti\inst{1}
          \and
          A. Zanella\inst{1}
          \and
          S. Tonnesen\inst{3}
          \and
          J. Fritz\inst{4}
          \and
          B. Vulcani\inst{1}
          \and
          C. Bacchini\inst{1}
          \and
          N. Akerman\inst{1,2}
          \and
          A. Kulier\inst{1}
          \and
          N. Tomicic\inst{5}
          \and
          R. Smith\inst{6}
          \and
          A. Wolter\inst{7}
          }

   \institute{INAF-Osservatorio Astronomico di Padova, Vicolo dell'Osservatorio 5, 35122 Padova, Italy
    \and
    Dipartimento di Fisica e Astronomia, Universit\`a di Padova, Vicolo Osservatorio 3, 35122 Padova, Italy
    \and
    Flatiron Institute, CCA, 162 5th Avenue, New York, NY 10010, USA
    \and
    Instituto de Radioastronom\'ia y Astrof\'isica, Universidad Nacional Aut\'onoma de M\'exico, Morelia, Michoac\'an, 58089 M\'exico
    \and
    Department of Physics, Faculty of Science, University of Zagreb, Bijeni\v{c}ka 32, 10 000 Zagreb, Croatia
    \and
    Departamento de Física, Universidad Técnica Federico Santa María, Avenida Vicuña Mackenna 3939, San Joaquín, Santiago, Chile
    \and
    INAF-Osservatorio Astronomico di Brera, via Brera 28, I-20121 Milano, Italy
    }

   \date{Received October 28, 1989; accepted March 16, 1997}
 
  \abstract
   {}
   {In this work, we aim to characterize the stellar populations of star-forming regions detached from the stellar disks of galaxies undergoing ram-pressure stripping.}
   {Using images collected with the WFC3 camera on board of the Hubble Space Telescope, we detect stellar clumps in continuum-subtracted $H\alpha$ and ultraviolet (F275W filter), such clumps are often embedded in larger regions (star-forming complexes) detected in the optical (F606W filter). 
   Our sample includes 347 $H\alpha$ clumps, 851 F275W clumps and 296 star-forming complexes.
   We model the photometry of these objects in 5 bands using {\sc bagpipes} to obtain their stellar population parameters.}
   {The median mass-weighted stellar ages are 27 Myr for $H\alpha$ clumps and 39 Myr for F275W clumps and star-forming complexes, but the oldest stars in the complexes can be older than $\sim$300 Myr which indicates that star-formation is sustained for long periods of time. 
   Stellar masses vary from 10$^{3.5}$ to 10$^{7.1}$ $M_\odot$, with star-forming complexes being more massive objects in the sample. 
   Clumps and complexes found further away from the host galaxy are on average younger, less massive and less obscured by dust. 
   We interpret these trends as due to the effect of ram-pressure in different phases of the interstellar medium.
   $H\alpha$ clumps form a well-defined sequence in the stellar mass--SFR plane with slope 0.73. Some F275W clumps and star-forming complexes follow the same sequence while others stray away from it and passively age. 
   The difference in mean stellar age between a complex and its youngest embedded clump scales with the distance between the clump and the center of the optical emission of the complex, with the most displaced clumps being hosted by the most elongated complexes.
   This is consistent with a fireball-like morphology, where star-formation proceeds in a small portion of the complex while older stars are left behind producing a linear stellar population gradient. 
   The stellar masses of star-forming complexes are consistent with the ones of globular clusters, but their stellar mass surface densities are lower by 2 dex, and their properties are more consistent with the population of dwarf galaxies in clusters.}
   {}


   \maketitle

\section{Introduction}\label{sec:intro}

The interaction with the hot gas of the intracluster medium (ICM) may remove gas from the interstellar medium (ISM) of galaxies falling into a galaxy cluster. 
This interaction is known as ram-pressure stripping \citep[RPS,][]{Gunn&Gott}. 
The most extreme examples of RPS are showcased by jellyfish galaxies, objects in the peak phase of stripping that display long tails of stripped gas.
Jellyfish galaxies can be identified in a variety of wavelengths, from X-rays to cold gas \citep{Poggianti2019JW100, Moretti2020, Roberts2021, Boselli2022, Rohr2023}.
In many cases, the tails of jellyfish galaxies can harbor star-forming regions \citep[see simulations by][]{Kapferer2009,Tonnesen2012} and are thus detectable in ionized gas and in the ultraviolet (UV) \citep[e.g][]{Fritz2017,George2018,George2023}. However, there is no consensus on how frequently star-formation is present in these tails, either in simulations \citep{Bekki2014, Steinhauser2016} or in observations \citep{Laudari2022}.

One of the largest samples of confirmed jellyfish galaxies is that of the GASP survey \citep[GAs Stripping Phenomena in galaxies with MUSE,][]{Poggianti2017}. GASP observed 114 galaxies with the MUSE Integral Field spectrograph on the VLT with the goal of clarifying how, where and when gas is removed from galaxies. 
The sample includes several galaxies that are undergoing RPS, some of which are textbook examples of jellyfish galaxies, while others have less prominent features.
The MUSE data provided by GASP reveal a clumpy structure in the $H\alpha$ images of the tails. Emission line diagnostic diagrams confirm these $H\alpha$ clumps as star-forming regions originating from the stripped gas, with 
median star-formation rates (SFRs) of $0.003 M_\odot/\mathrm{yr}$, compatible with the low end of the $\mathrm{SFR}$ distribution for $H\alpha$ clumps in the disks of the same galaxies. Their stellar masses vary typically between $\sim 10^{4}$ and $\sim 10^{7}\,M_\odot$, which is comparable to globular clusters and dwarf galaxies \citep{Poggianti2019, Vulcani2020b}.
The $H\alpha$ clumps in ram-pressure stripped tails observed by GASP are surrounded by a diffuse ionized gas (DIG) component where ionization is at least partially powered by a non-stellar source, possibly the mixing of the ISM and ICM gas \citep{Neven2021}.






High spatial resolution observations of nearby galaxies using the Hubble Space Telescope (HST) such as the ones provided by The Legacy ExtraGalactic UV Survey \citep[LEGUS][]{Calzetti2015} and the Physics at High Angular resolution in Nearby GalaxieS \citep[PHANGS][]{Schinnerer2019} projects
allow for very detailed studies of the structure of extragalactic star-forming regions that cannot be performed with ground-based instruments such as MUSE.
For example, these studies have shown that star-formation is a hierarchical process \citep{Elmegreen2006, Gouliermis2015, Gouliermis2017} in both space and time, with young structures being highly clustered and becoming less spatially correlated as they age \citep[e.g][]{Grasha2017a, Grasha2017b}.
High resolution studies using HST have also been performed for galaxies undergoing RPS, most notably in the Coma and Virgo clusters. These works highlight the decoupling of dense gas from the surrounding ISM in the disk \citep[][]{Abramson2014, Kenney2015} and constrain the sizes and stellar populations of stellar sources in the stripped tails \citep[e.g][]{Cramer2019, Waldron2023}. 

Recently, \cite{Marco2023} have obtained UVIS/HST images of 6 jellyfish galaxies from the GASP survey, allowing a statistical study of star-forming regions in the tails and disks of these objects at sub-kpc resolution.
The observations were made in 5 filters, from the UV to the I-band, including a narrow-band filter around the $H\alpha$ emission line.
\cite{Eric2023} used the \textsc{astrodendro} software to identify star-forming clumps in the $H\alpha$ and F275W images of these galaxies; 
these clumps are shown to be embedded in
 larger structures detected in the F606W filter, called star-forming complexes.
The authors find that these clumps have enhanced $H\alpha$ luminosity at a given size, falling in a luminosity-size relation that is more similar to the one observed for starburst galaxies than that of normal star-forming galaxies.
In the following paper of the series \citep{Giunchi2023b}, the authors focus on the morphological properties of the clumps and complexes, finding that they have slightly elongated structures, with
clumps being displaced by 0.1 to 1 kpc from the center of the star-forming complex on which they are embedded. These results, confirmed by the visual inspection of the images, show that these objects are organized in structures known as ``fireballs" \citep{Cortese2007, Yoshida2008, Kenney2014, Jachym2017, Waldron2023}.

In this work, we model the spectral energy distribution (SED) of the clumps detected by \cite{Eric2023} in order to determine their ages, masses, star-formation rates and other properties, extending previous studies of stellar populations of clumps in these galaxies to the sub-kpc regime.
The data-set is presented in section \ref{sec:data} and the SED fitting procedure is described in section \ref{sec:model}.
In sections \ref{sec:results} and \ref{sec:complexes}, we present results of the SED fitting and explore applications to several science cases.

We assume a standard $\Lambda$CDM cosmology with $\Omega_{\rm M}=0.3$, $\Omega_\Lambda=0.7$ and $h=0.7$. The chosen epoch for right ascension and declination (RA and Dec) is J2000.

\section{Observations and clump detection} \label{sec:data}

This paper relies on data from \cite{Marco2023} and the clump detection method of \cite{Eric2023}. All procedures are described in detail in the respective papers, but we 
 summarize the main aspects of the data and methodology in this section.

\subsection{The data}

 
Our work is based on images obtained using the UVIS channel of the WFC3 camera on board of HST for six jellyfish galaxies in the GASP sample: JO175, JO201, JO204, JO206, JW39 and JW100.

These galaxies were observed in five filters: F275W, F336W, F606W, F680N and F814W, sampling the wavelength range from $\sim 2475\,\mathrm{\AA}$ to $\sim9560\,\mathrm{\AA}$, with F680N covering the $H\alpha$ emission line in the redshift range of the targeted galaxies (0.042 - 0.066). 
The FWHM of the PSF of the observations in this redshift range corresponds to $\sim 70$ pc.


The data are available in the Mikulski Archive for Space Telescopes (MAST): \href{10.17909/tms2-9250}{10.17909/tms2-9250}. The ID of the project is GO-16223; PI Gullieuszik.

\subsection{Clump detection}\label{sec:detection}

Clump candidates are detected using \textsc{Astrodendro}\footnote{\url{https://dendrograms.readthedocs.io/en/stable/index.html}}, a Python package designed to define bright regions and brighter sub-regions inside them, building a tree structure formed by trunks (the ground levels), branches (the intermediate levels) and leaves (the top levels, containing no sub-regions). 

The clumps analyzed in this work are located in two sub-regions of the galaxies: (i) extraplanar, the region where clumps show clear signs of disturbance due to ram-pressure but still overlap (at least in projection) with the outskirts of the stellar disk (as defined by \citealt{Eric2023}); and (ii) tails, where clumps do not overlap with the disks.

\textsc{Astrodendro} was run independently in the H$\alpha$ (continuum subtracted) and F275W images, set to detect structures where at least 5 adjacent pixels have fluxes 2$\sigma$ above the background level in the detection filter. 
To avoid spurious detection due to background objects, in the tails we keep only candidates either (i) matched with MUSE clumps \citep{Poggianti2019}; (ii) with a redshift consistent with the one of the galaxy in the corresponding region of its MUSE data cube or (iii) emitting in all the filters and with a positive H$\alpha$ flux.
We refer to \cite{Eric2023} for a flowchart that illustrates the aforementioned selection criteria.
To ensure that the physical properties derived from the SED fitting can be adequately constrained, we further select only \textsc{Astrodendro} trunks and leaves that have a signal-to-noise ratio SNR$>2$ in all of the 5 filters; these are only the criteria in our selection that differ from \cite{Eric2023}.
This leads to an initial sample of 409 H$\alpha$ clumps and 1208 F275W clumps (combining tail and extraplanar regions).

In order to trace the emission of older stellar population and recover the whole stellar mass formed from stripped gas in the tails, we ran \textsc{Astrodendro} also on the F606W filter.
Throughout the paper, these larger structures detected in the F606W filter are referred to as star-forming complexes, or just ``complexes''.
In order to avoid the diffuse emission of the galactic stellar disk, which is very bright in the F606W filter,
the detection of these star-forming complexes is performed only in the tails (not in extraplanar regions), and the detection threshold was increased to 3$\sigma$. Further selection criteria require the candidate to be an \textsc{Astrodendro} trunk matched to at least one F275W or H$\alpha$ clump\footnote{By ``matched" we mean that the complex overlaps with the clump by at least one pixel.}. The total number of star-forming complexes is 338.

From these samples, we select a sub-sample of spatially resolved clumps and complexes. The size of a clump is defined as $2 R_C$, where $R_C$ is the PSF-corrected core radius, i.e. the geometric mean of the semi-major and semi-minor axes of the clumps, computed as the standard deviations of the clump surface brightness distribution along and perpendicular to the direction of maximum elongation. We define a clump or a complex as resolved if its size is larger than $2\mathrm{FWHM_{PSF}}$ ($\sim140$ pc). The median sizes of resolved H$\alpha$ clumps, F275W clumps and star-forming complexes are $\sim 201$ pc, $\sim 214$ pc and $\sim 251$ pc, respectively.
Typically, $15-20\%$ of $H\alpha$ and F275W clumps and about half of the star-forming complexes are resolved.
Throughout the paper, morphological quantities such as size, axial ratio and stellar mass surface density are reported only for these resolved objects.

We anticipate that some of the clumps and complexes that were detected will not be used in this work, as some quality control cuts are applied based on the results of the stellar population modeling. These additional cuts are discussed in sec. \ref{sec:final_sample}, where we present our final sample.

\section{Deriving physical properties}\label{sec:model}

We model the stellar populations of clumps and complexes in our sample by fitting all 5 observed fluxes with the SED fitting code {\sc bagpipes} \cite[Bayesian Analysis of Galaxies for Physical Inference and Parameter EStimation,][]{BAGPIPES}.
For each band, the fitted fluxes are measured as the sum of the fluxes of all pixels belonging to a given clump or complex as defined by {\sc astrodendro}.
We adopt the 2016 update of the simple stellar population (SSP) models from \cite{Bruzual2003} with a \cite{Kroupa2001} initial mass function; the solar metallicity in these models is $Z_\odot = 0.017$. 
{\sc bagpipes} returns posterior probability distributions (PDFs) for each of the measured properties; throughout the paper we use the median of the PDFs when referring to the value of a physical property of an object.

\subsection{Model description} \label{sec:synthesis}

Star-formation histories (SFHs) of clumps and star-forming complexes are parameterized as a single delayed exponential \citep[see][]{Carnall2019} following
\begin{equation}
\textsc{SFR}(t) \propto
    \begin{cases}
    (t-t_0) e^{-\frac{t-t_0}{\tau}} & t>t_0 \\
    0 & t<t_0.
    \end{cases}
    \label{eq:sfh}
\end{equation}
where $t$ is the time since the Big Bang, $t_0$ the time when star-formation starts and $\tau$ is the timescale of the decline of star-formation. This parametrization can be quite flexible as it reduces to a single burst for $\tau \sim 0$.

We have also tested two other models for the SFHs: (i) instantaneous bursts (or single SSPs) and (ii) square bursts (constant star-formation during a certain period of time). In general, both models perform somewhat similarly to the delayed exponential. However, single SSPs are not flexible enough to reproduce some of the observed colors (see section \ref{sec:t0_tau}). Square bursts yield very similar fits to delayed exponentials, with just slightly higher residuals.

\begin{figure*}[ht!]
\includegraphics[width=\textwidth]{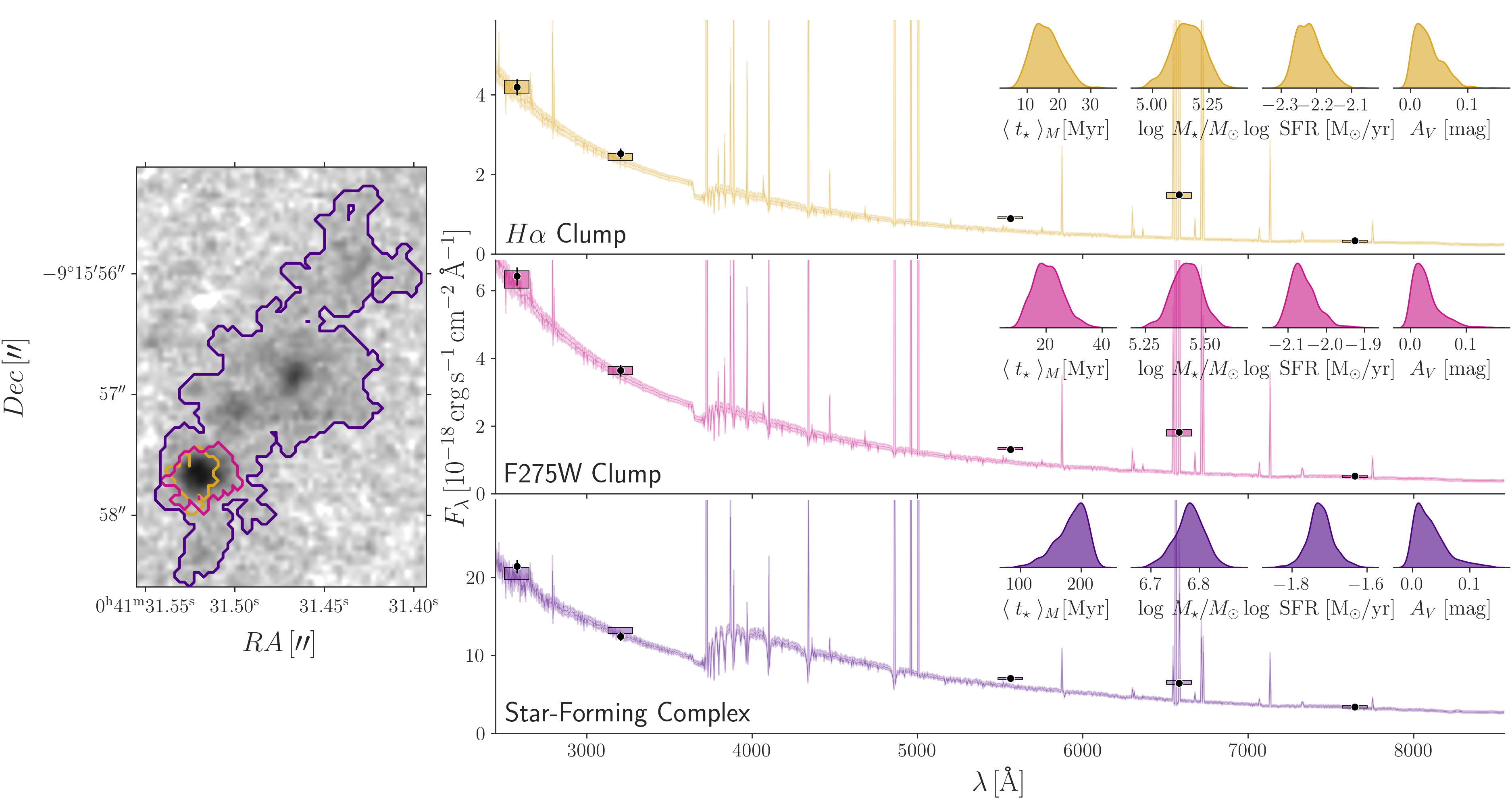}
\caption{Examples of \textsc{bagpipes} fits for clumps selected from H$\alpha$, F275W and F606W images. On the left: F275W image of a region in the tail of JO201; regions in golden yellow, pink and purple indicate an H$\alpha$ clump, a UV clump and a star-forming complex, respectively. On the right: Examples of \textsc{bagpipes} fits to the photometry of the regions indicated on the image to the left; shaded region corresponds to the full range of the model spectra PDF and are plotted with the same color as the corresponding region in the image;
black points with errorbars (generally too small to be seen) indicate the observed photometric fluxes in the 5 HST bands used in this work. Rectangles show the region between the 1\% and 99\% percentiles of the photometric fluxes fitted by \textsc{bagpipes}. 
Inset plots in the right panels show the posterior PDFs of mass-weighted ages ($\langle t_\star \rangle$), stellar masses ($\log\,M_\star/M_\odot)$, star-formation rates (SFRs) and dust attenuation ($A_V$) derived for each object.}
\label{fig:example}
\end{figure*}

Dust is modeled using the Milky Way extinction curve from \cite{CCM} with $R_V=3.1$ and two foreground dust screens with different $V$-band extinction $A_V$, for stellar populations older and younger than 20 Myr.  
We have also experimented with the possibility of setting the $A_V$ priors according to the Balmer decrement of star-forming regions detected in $H\alpha$ maps obtained from the MUSE data cubes available for these galaxies. However, for $H\alpha$ clumps we found that better fits are usually achieved with slightly lower $A_V$ ($\sim0.3$ mag on average) and due to the substructure of HST clumps the same idea doesn't apply for F275W clumps and star-forming complexes, which is why we chose to keep $A_V$ as a free parameter.

Emission lines and nebular continuum emission (derived using {\sc cloudy} \cite{Cloudy}) are included in the models for stellar populations as old as 20 Myr. 

Priors on the parameters fitted by {\sc bagpipes} are set according to the expected properties for these recently formed objects. 
Ages are allowed to vary between 0 and 500 Myr and the total mass assembled into stars is left to vary from 0 to $10^{10}M_\odot$.
Stellar metallicity follows a Gaussian prior with mean $\mu=Z_\odot$ and standard deviation $\sigma=0.25\, Z_\odot$.
The $\tau$ parameter in the delayed exponential varies from 10$^{-3}$ to 500 Myr according to a logarithmic prior; this prior was adopted to allow $\tau$ to converge to very low values when necessary, eventually reducing the delayed exponential to a single burst of star-formation.
$A_V$ varies from 0 to 1.25 magnitudes and the ratio between the $A_V$ of the young and the old populations varies from 1 and 2.5. The ionization parameter $\log\,U$ varies from -3.5 to -2 and gas metallicity is the same as the stellar metallicity. Redshift is fixed at the galaxy redshift measured from the MUSE datacubes \citep{Marco2020}, as variations within a single galaxy have a negligible effect on the photometry.

Examples of {\sc bagpipes} fits for a star-forming complex and two embedded clumps (one detected in $H\alpha$ and one in F275W) are shown in Fig. \ref{fig:example}. Corner plots showing the covariance between the main derived parameters are shown in appendix \ref{sec:corners}.

\subsection{Quality control cuts and final sample}\label{sec:final_sample}

To evaluate how the underlying stellar populations in the galaxy disk might affect our results,
we have tested a star-formation history with two components: one with the same configuration described above, and an older one representing the underlying stellar population of the disk. 
This is expected to be an issue especially in the extraplanar regions (which overlap with the disk), but also for tail clumps near the stellar disk contour delimited by \citet{Eric2023}. Star-forming complexes are less prone to this effect as they are detected in the F606W filter and by selection are well detached from the disk.
This test showed that the stellar mass contribution of the old components was generally very low ($\lesssim1\%$) and the properties (ages, dust content, stellar masses and $\mathrm{SFRs}$) derived for the clumps were unaffected. 
However, in some cases this experiment has revealed non-negligible old components, which leads to unconstrained ages ($\sim$flat PDFs) for the young components. 
This happens for 29 $H\alpha$ clumps (7.0\%) and 204 F275W clumps (15.7\%), all of which are removed from the final sample.
Of these, 21 $H\alpha$ clumps and 123 F275W clumps are in the extraplanar region of JW100.

In some cases, the fluxes of our models are outside the 2$\sigma$ error bars of the observations for one or more filters, thus the fit cannot be considered satisfactory. This happens mostly for clumps detected in F275W: due to the slightly lower signal to noise of this filter, the detection sometimes misses some of the faint pixels that are below the background level in F275W but still contribute to F336W. In these cases, the F275W flux tends to be slightly under predicted, and F336W slightly over predicted.
To ensure data quality, we have discarded objects for which either the median of the model flux PDF is outside error bars of the observed photometry for more than one filter or the interquartile region of the model flux PDF does not intersect with the observed flux error bar for one of the filters.
These criteria further remove 33 (8.6\%) $H\alpha$ clumps, 243 (22.2\%) F275W clumps and 42 (12\%) star-forming complexes. 
Examples of fits that failed the quality control are provided in Appendix \ref{sec:bad_fits}.

After these cuts, our final sample contains 347 $H\alpha$ clumps, 851 F275W clumps and 296 star-forming complexes. The final number of clumps and complexes in each galaxy and the number of resolved objects according to the criterion described in section \ref{sec:detection} are shown in table \ref{tab:sample}. For information on our galaxy sample and their host clusters, we refer to \cite{Eric2023}.



\begin{table}
\centering
\caption{Number of objects in each galaxy (and in total) that are included in our final sample.}
\begin{tabular}{llll}
\hline
       & \multicolumn{3}{c}{Number of objects (all/resolved)}                    \\ \hline 
Galaxy & $H\alpha$ Clumps & F275W Clumps & Complexes \\ \hline
JO175  & 30/1             & 55/7         & 21/11                     \\
JO201  & 86/14            & 307/65       & 85/49                     \\
JO204  & 28/4             & 96/14       & 45/24                     \\
JO206  & 123/17           & 196/33       & 59/29                     \\
JW39   & 22/2             & 70/4         & 53/22                     \\
JW100  & 58/8             & 127/19       & 33/22                   \\
Total  & 347/46           & 851/142      & 296/157                   \\
\hline
\end{tabular}
\label{tab:sample}
\end{table}

\subsection{$t_0^\star/\tau$ and SFH complexity}\label{sec:t0_tau}

\begin{figure}[ht]
\includegraphics[width=\columnwidth]{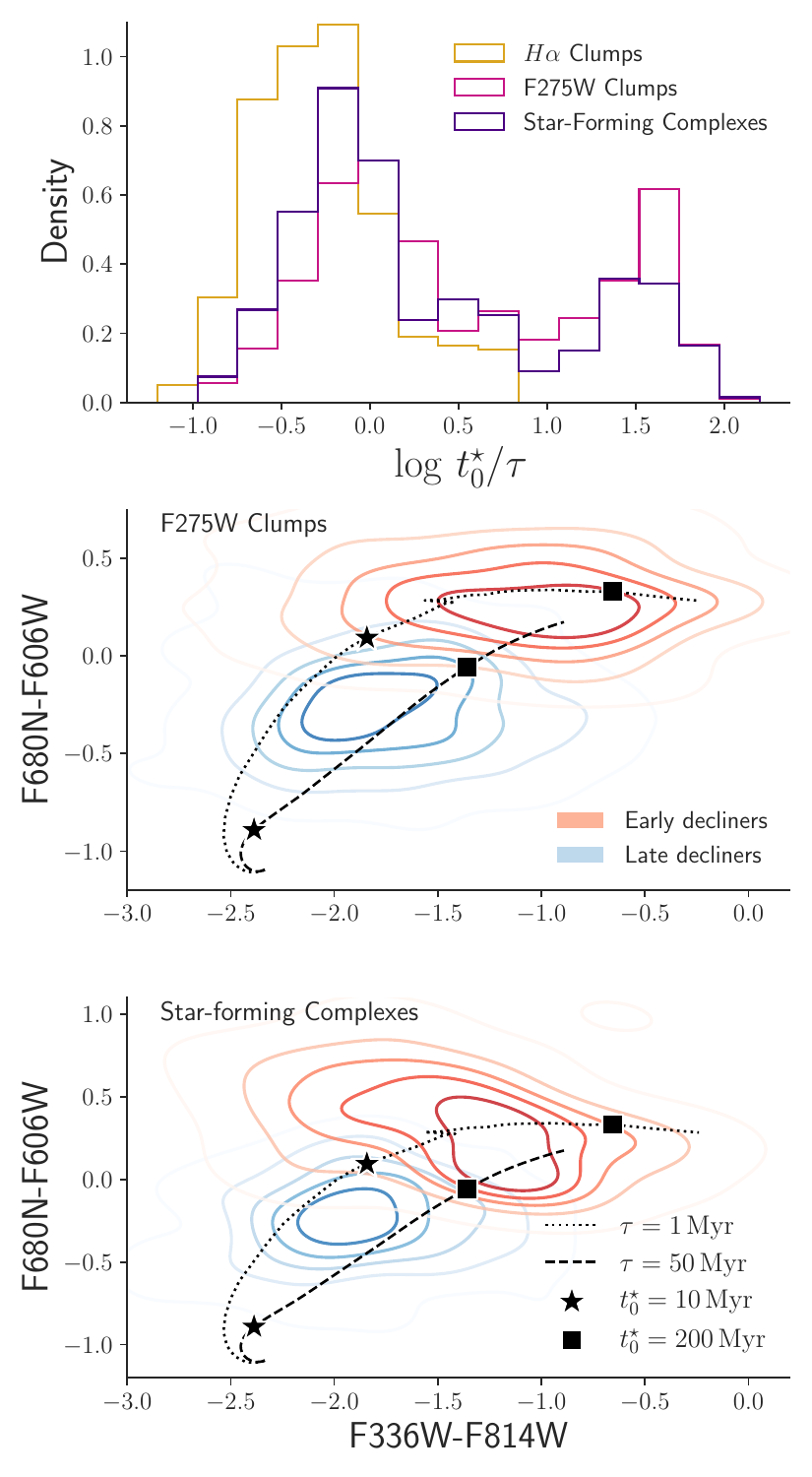}
\caption{Top: histograms tracing the distributions of the SFH decay proxy $t_0^\star/\tau$ for $H\alpha$ clumps (gold), F275W clumps (pink) and star-forming complexes (purple). Middle and bottom panels show F680N-F606W vs. F336W-F814W color-color diagrams for F275W clumps and star-forming complexes, respectively.
Contours in the color-color diagrams trace 2D kernel density estimations tracing the loci of late (blue) and early (red) decliners (see text for details); dotted and dashed lines indicate {\sc bagpipes} models of different ages and $\tau=1 \, \mathrm{Myr}$ and $\mathrm{50\,Myr}$, respectively. The position of models with $t_0^\star$ values of $10\,\mathrm{Myr}$ (stars) and $200 \, \mathrm{Myr}$ (squares) along each of the lines of constant $\tau$ is marked in both diagrams.}
\label{fig:t0_tau}
\end{figure}

As mentioned in sec. \ref{sec:synthesis}, our delayed exponential model can reproduce both burst-like and slowly decaying SFHs, but it is worthwhile (and wise) to ask if such complexity is needed to reproduce the observations.
To verify that, we use the ratio between the age of the oldest stars in the model ($t_0^\star$\footnote{$t_0^\star$ corresponds to the time spanned between $t_0$ in equation \ref{eq:sfh} and the age of the universe at the observed redshift.}) and $\tau$ (see equation \ref{eq:sfh}) as a normalized proxy of the SFR evolution, as this ratio effectively measures how much the SFR has declined from its maximum value. Histograms of $t_0^\star/\tau$ for our sample of clumps and complexes are shown in the top panel of Fig. \ref{fig:t0_tau}. $H\alpha$ clumps have typically low $t_0^\star/\tau$ indicative of SFHs that have not declined significantly (consistent with constant SFR), while F275W clumps and star-forming complexes show bimodal distributions; we refer to these two distinct populations as early (high $t_0^\star/\tau$) and late (low $t_0^\star/\tau$) decliners.
To separate these classes, we have used a Gaussian mixture model with three clusters: one for each population and a third one to account for intermediate objects. The model was trained in the combined sample of clumps and complexes and then applied for the classification of each sub-sample.
We classify 90\% of $H\alpha$ clumps, 49\% of F275W clumps and 61\% of star-forming complexes as late decliners, while 30\% of F275W clumps, 22\% of star-forming complexes and none of the $H\alpha$ clumps are early decliners according to the Gaussian Mixture model.

We found that early and late decliners occupy well-defined loci in color-color diagrams, and are especially well separated in F680N-F606W, which is a color that traces the strength of $H\alpha$ emission compared to the underlying continuum (similarly to an equivalent width, with stronger $H\alpha$ corresponding to bluer color). In the central and bottom panels of Fig. \ref{fig:t0_tau} we show the F680N-F606W vs. F336W-F814W color-color diagram for F275W clumps and star-forming complexes, respectively; with contours tracing the distribution of fast and slow decliners.
The population of clumps and complexes that is intermediate in $t_0^\star/\tau$ is also intermediate in the color-color diagrams, but was omitted from the figure to improve visualization.
We compare the colors of early and late decliners with two model tracks generated with {\sc bagpipes} (according to eq. \ref{eq:sfh}) for ages between 1 and 300 Myr and $\tau$ between 1 and 50 Myr. Other parameters are kept constant: $Z=Z_\odot$, $\log\,U=-2.75$, and $A_V=0.1$ for stars of all ages.

The two model tracks start indistinguishable from each other at $t_0^\star=1\,\mathrm{Myr}$, but quickly diverge to different trajectories in the diagram. 
The $\tau=1\,\mathrm{Myr}$ models cross the threshold between late and early decliners at $t_0^\star\sim10\,\mathrm{Myr}$ (corresponding to $\log\,t_0^\star/\tau \sim 1$) and evolve to redder F336W-F814W with little change in F680N-F606W after that.
On the other hand, models with $\tau=50\,\mathrm{Myr}$ 
retain blue F680N-F606W for a long time (up to $t_0^\star\sim200\,\mathrm{Myr}$) because they keep generating strong $H\alpha$ emission as they become older and their broad-band colors redden.

This indicates that some level of complexity in the SFH models is indeed required to reproduce the mixture of $H\alpha$-strong and $H\alpha$-weak objects with a variety of broad-band colors.
We also note that although our analysis is able to distinguish between two broad ranges of $\log\,t_0^\star/\tau$ and we show why these two regimes are necessary to reproduce the data, we lack the constraining power to obtain precise estimates for this parameter.

\section{Properties of clumps and complexes}\label{sec:results}

The main quantities of interest for this paper are the stellar ages, 
stellar masses, star-formation rates (calculated in the past $10\,\mathrm{Myr}$) and dust attenuation of the clumps and complexes. Here and throughout most of the paper we use mass-weighted stellar ages ($\langle\,t_\star\rangle_M$). These are very similar to luminosity-weighted ages for these objects as the mass-to-light ratios do not vary widely in these age ranges (see \citealt{Cid2005} for a discussion on age definitions). 
The reported $A_V$ values correspond to the attenuation affecting stars younger than 20 Myr, and to ensure meaningful measurements we analyze this quantity only for objects classified as late decliners in sec. \ref{sec:t0_tau}, i.e objects with significant amounts of young (< 20 Myr) stars. 

\subsection{Distributions of measured properties}\label{dists}

\begin{figure*}[ht]
\includegraphics[width=\textwidth]{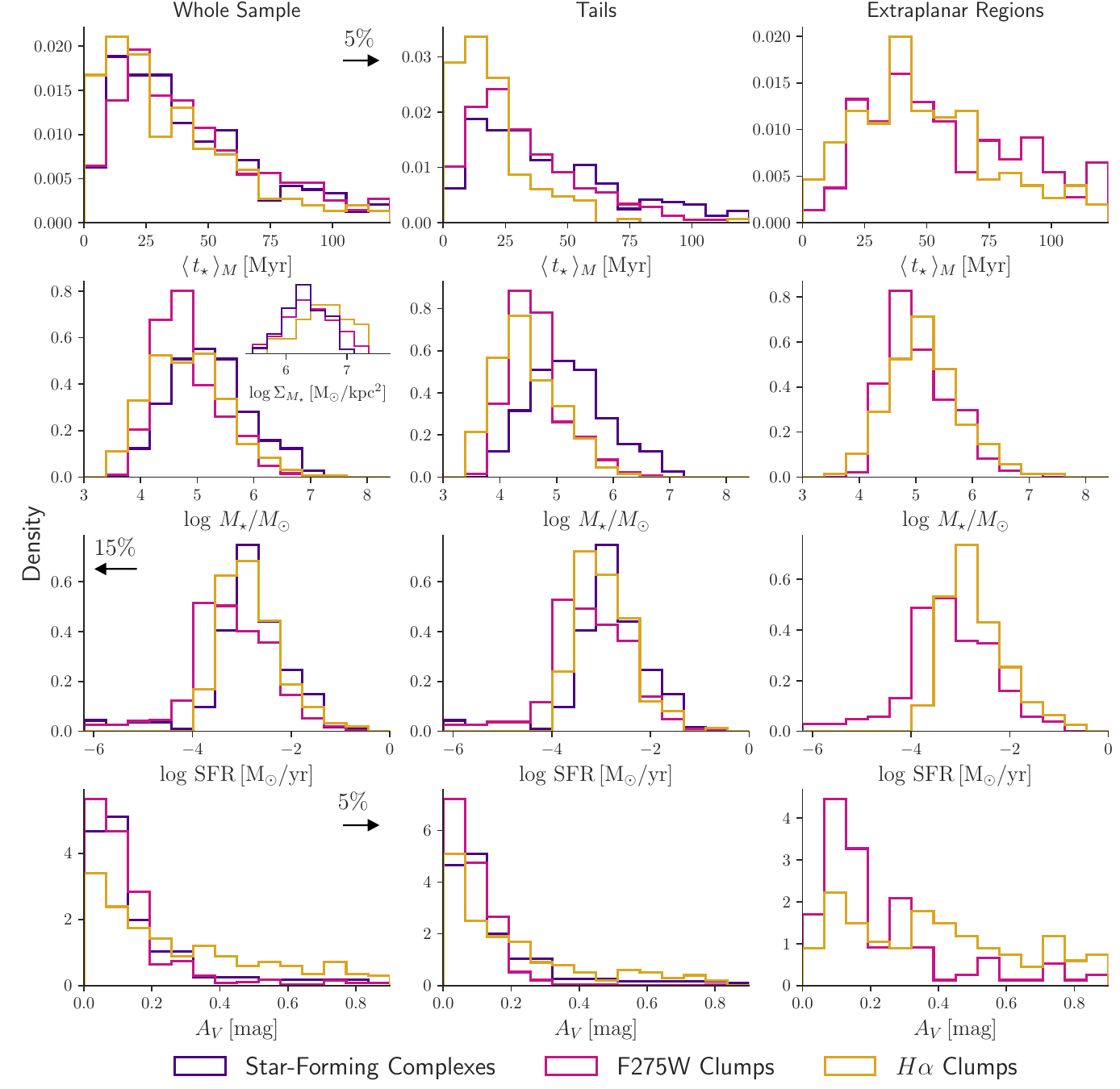}
\caption{Distributions of (from top to bottom) mass-weighted ages, stellar masses, star-formation rates and dust attenuation of clumps detected in $H\alpha$ (in gold) and F275W (in dark pink), and star-forming complexes (in purple). 
We limit the histograms to the 95th percentiles of $\langle\,t_\star\rangle_M$ and $A_V$, and to the 15th percentile of SFR. These percentiles are calculated for the combined (all clumps and complexes) samples and the ranges are kept the same in all histograms of the same variable.
Left panels show the general samples, tail clumps and complexes are in the middle panels and extraplanar clumps are on the right. Histograms are normalized to have the same area.}
\label{fig:all_hist}
\end{figure*}

The distributions of $\langle\,t_\star\rangle_M$, $\log\,M_\star$, SFR and $A_V$ for our sample of clumps and complexes in tails and extraplanar regions are shown in Fig. \ref{fig:all_hist}. 

\begin{figure*}[ht]
\includegraphics[width=\textwidth]{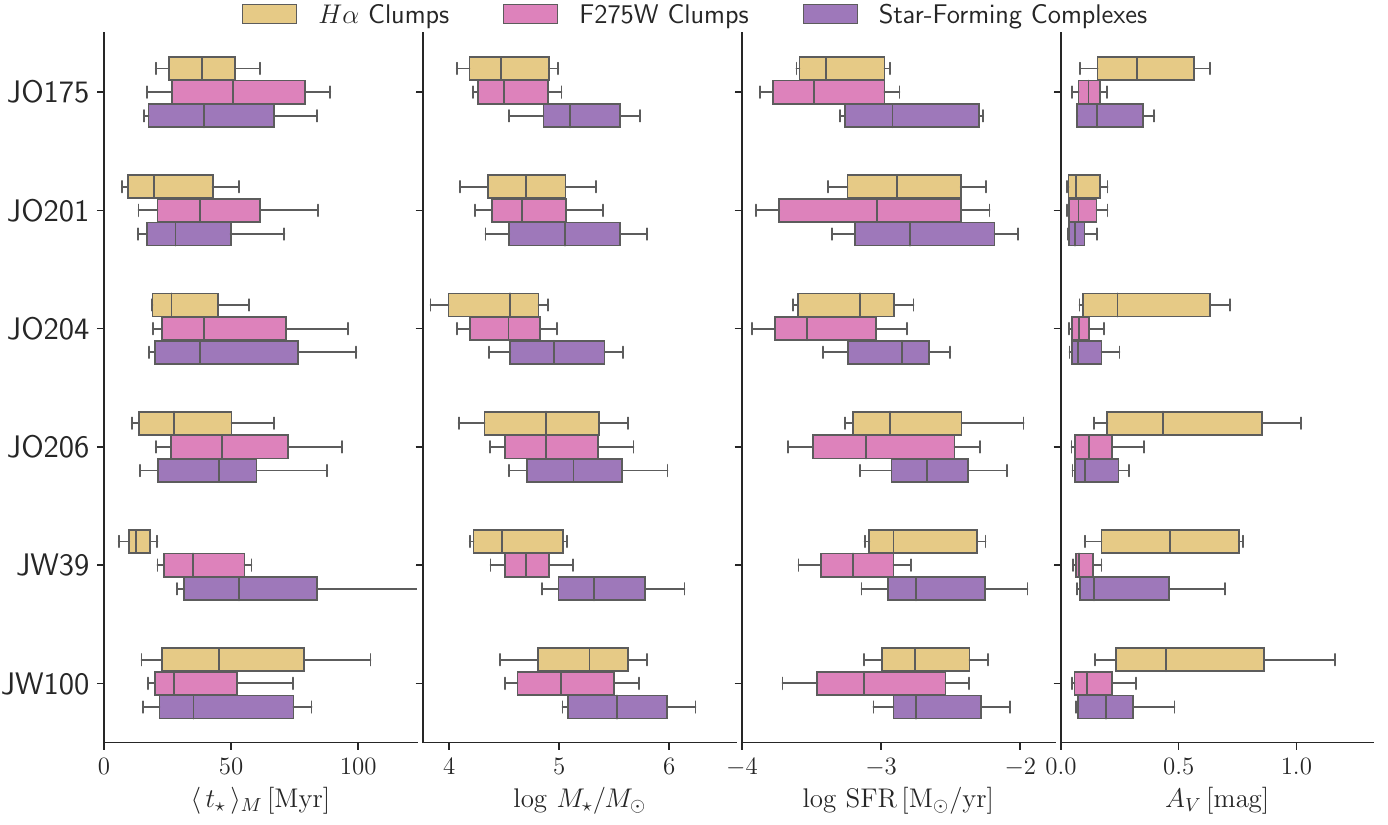}
\caption{Box plots showing galaxy-to-galaxy variations of the distributions shown in Fig. \ref{fig:all_hist}. As before, clumps detected in $H\alpha$ and F275W are shown in gold and in dark pink, respectively, and star-forming complexes are shown in purple. Boxes span the interquartile regions, horizontal lines in the boxes indicate the median and whiskers mark the 15\% and 85\% percentiles of the distributions.}
\label{fig:all_galbygal}
\end{figure*}

As expected, $H\alpha$-detected clumps are the youngest, as these regions contain mostly very young stars still enclosed in their birth clouds. Considering the whole sample of $H\alpha$ clumps, the median mass-weighted age is 27 Myrs. The ages vary between sub-samples: tail clumps have a median mass-weighted age of 16 Myrs, while extraplanar clumps are relatively older (median $\langle\,t_\star\,\rangle_M=46$ Myrs).
In comparison, clumps detected in F275W are slightly older, with median $\langle\,t_\star\,\rangle_M$ of 39 Myrs for the whole sample, 27 Myrs in tails and 58 Myrs in the extraplanar regions. 
This confirms the observational interpretation that these regions include stars that have just left their birth clouds and have slightly older ages than the ones in $H\alpha$-emitting regions.
Star-forming complexes also have a median $\langle\,t_\star\,\rangle_M$ of 39 Myrs, but we remind the reader that these objects are only detected in the tails, where both clump samples are younger. 
When masking out the $H\alpha$ and F275W clumps inside the complex and leaving only the flux of the region detected in F606W, we obtain a much older median $\langle\,t_\star\,\rangle_M$ of 84 Myrs.
This shows that the diffuse emission detected in the F606W filter is really associated with an older stellar population, and not just a fainter region of the F275W clump that is not detected due to the lower SNR of this filter.

The median stellar mass of $H\alpha$ clumps is $10^{4.8} M_\odot$ for the whole sample, $10^{4.4} M_\odot$ in the tails and $10^{5.1} M_\odot$ in extraplanar regions.   
For F275W clumps the median stellar masses are very similar: $10^{4.7} M_\odot$ for the whole sample, $10^{4.4} M_\odot$ for tail clumps and $10^{5} M_\odot$ for extraplanar clumps. On the other hand, star-forming complexes are more massive, with median $\log M/M_\odot = 5.2$.
The full stellar mass range of our sample is from $10^{3.5}$ to $10^{7.1}\;M_\odot$, similar to the one of globular clusters and dwarf galaxies. 

Recent works in the literature also derive ages and masses for stellar clumps in the tails of jellyfish galaxies based on HST photometry. These works report stellar masses similar to the ones we obtain, but ages that are relatively younger: 1 -- 35 Myr for D100 \citep{Cramer2019} and < 10 Myr for ESO 137-001 \citep{Waldron2023}. Although these are different galaxies and the values might indeed differ, differences in methodology and data certainly play a role in producing this discrepancy. 
Both \cite{Cramer2019} and \cite{Waldron2023} use Starburst99 and Cloudy to model colors, magnitudes and/or the equivalent width of H$\alpha$ of the clumps. These models are based on single bursts and therefore not comparable to the model used in this work. When limiting our SFHs to a single burst, we obtain median ages of 6 Myr to H$\alpha$ clumps and 15 Myr for F275W clumps, which is closer to the literature values.

Another important quantity for understanding the structure of star-forming regions is the average stellar mass surface density $\Sigma_{M_\star}$. 
With the diffraction limited images provided by HST, we are able to constrain $\Sigma_{M_\star}$ for a significant fraction of clumps and complexes, although many remain unresolved (see sec. \ref{sec:detection}). We note that in this work $\Sigma_{M_\star}$ is obtained by dividing the stellar mass of the object by its total physical area (the sum of the area of all pixels, converted to kpc$^2$ according to the galaxy redshift), which is slightly different from what is done in some works in the literature where $\Sigma_{M_\star}$ is calculated within an effective radius.
Histograms of $\log\,\Sigma_{M_\star}$ are shown as an inset in the stellar mass histogram for the whole sample. $H\alpha$ clumps are denser in stellar mass than F275W clumps, which in turn are denser than star-forming complexes, even though the median values are very similar: $\log\,\Sigma_{M_\star}/M_\odot\mathrm{kpc^{-2}}$ = 6.6 for $H\alpha$ clumps, 6.4 for F275W clumps and 6.3 for star-forming complexes.

As expected, $H\alpha$ clumps represent regions of ongoing star-formation and thus are all characterized by significant $\mathrm{SFRs}$ (median $\mathrm{SFR}$ of $10^{-2.9}\mathrm{M_\odot/yr}$ for the whole sample,  $10^{-3}\mathrm{M_\odot/yr}$ for tail clumps and  $10^{-2.8}\mathrm{M_\odot/yr}$ for extraplanar clumps).  
For F275W clumps and star-forming complexes, the distributions of $\log\,\mathrm{SFR}$ show a tail towards low values ($\log \, \mathrm{SFR}/\mathrm{M_\odot yr^{-1}} < -4.5$), these tails contain 26\% of F275W clumps and 18\% of star-forming complexes. This happens because some of these objects do not display any star-formation in the past 10 Myrs, but since we used a delayed exponential model for the SFH, we still measure $\mathrm{SFR}>0$, although very low. 
When considering only star-forming complexes with $\mathrm{SFR}>10^{-4.5}\mathrm{M_\odot/yr}$ (i.e neglecting the tail of the distribution) the median $\mathrm{SFR}$ is $10^{-2.75} \mathrm{M_\odot/yr}$, very similar to (and slightly higher than) the one of $H\alpha$ clumps in the tails, which can be understood as star-forming complexes may contain one or more $H\alpha$ clump(s). 
On the other hand, F275W clumps have lower SFRs when compared to $H\alpha$ clumps or star-forming complexes even when neglecting the tails of the distribution, their median $\mathrm{SFR}$ for clumps with $\mathrm{SFR}>10^{-4.5}\,\mathrm{M_\odot/yr}$ is $\sim10^{-3.15}\mathrm{M_\odot/yr}$ for both subsamples.
We note that a substantial fraction of the objects in our sample have $\mathrm{SFR}<10^{-3}\,M_\odot/yr$, which corresponds to a mass formed in the past $10\,\mathrm{Myr}$ that is smaller than $10^4\,M_\odot$, which is the typical value required for a fully sampled IMF \citep{daSilva2012}; despite this relevant caveat, we find no indication that objects with $\mathrm{SFRs}$ below this threshold behave in any peculiar way.
The $\mathrm{SFRs}$ of objects in our sample will be further explored in sec. \ref{sec:sfr-mass}.

When considering the whole sample or the tail regions,
the distribution of $A_V$ for $H\alpha$ clumps is slightly skewed towards larger values when compared to F275W clumps. However, all distributions are concentrated at low $A_V$, except for the population of $H\alpha$ clumps in extraplanar regions. Furthermore, F275W clumps in extraplanar regions are slightly more attenuated than their tail counterparts.
We do expect $H\alpha$ clumps to be more strongly attenuated by dust, in the same way that star-forming regions in galaxy disks are more attenuated than the older stellar populations \citep{Charlot2000}.
What is unclear is why the most attenuated clumps are in extraplanar regions. We see two possible explanations for this: (i) the extra $A_V$ might actually be due to diffuse (possibly stripped) dust in and around the galaxy disk; (ii) alternatively, this might be a consequence of the fact that extraplanar clumps are typically the oldest and thus had more time to form dust grains in-situ.
The two explanations are not mutually exclusive, and both effects might contribute to the observed $A_V$ differences.
However, both of these interpretations would point to the conclusion that dust has not been stripped out to large distances.

The distributions presented in Fig. \ref{fig:all_hist} remain fairly similar in all galaxies of our sample. Box plots showing the medians, interquartile regions and 15\% and 85\% percentiles of the distributions in individual galaxies are shown in Fig. \ref{fig:all_galbygal}. The distinction between tail and extraplanar clumps is neglected for simplicity, but also because some galaxies do not have enough clumps to produce meaningful distributions when making the distinction.

\subsection{Stellar Mass - SFR relation}\label{sec:sfr-mass}

\begin{figure}[ht]
\includegraphics[width=\columnwidth]{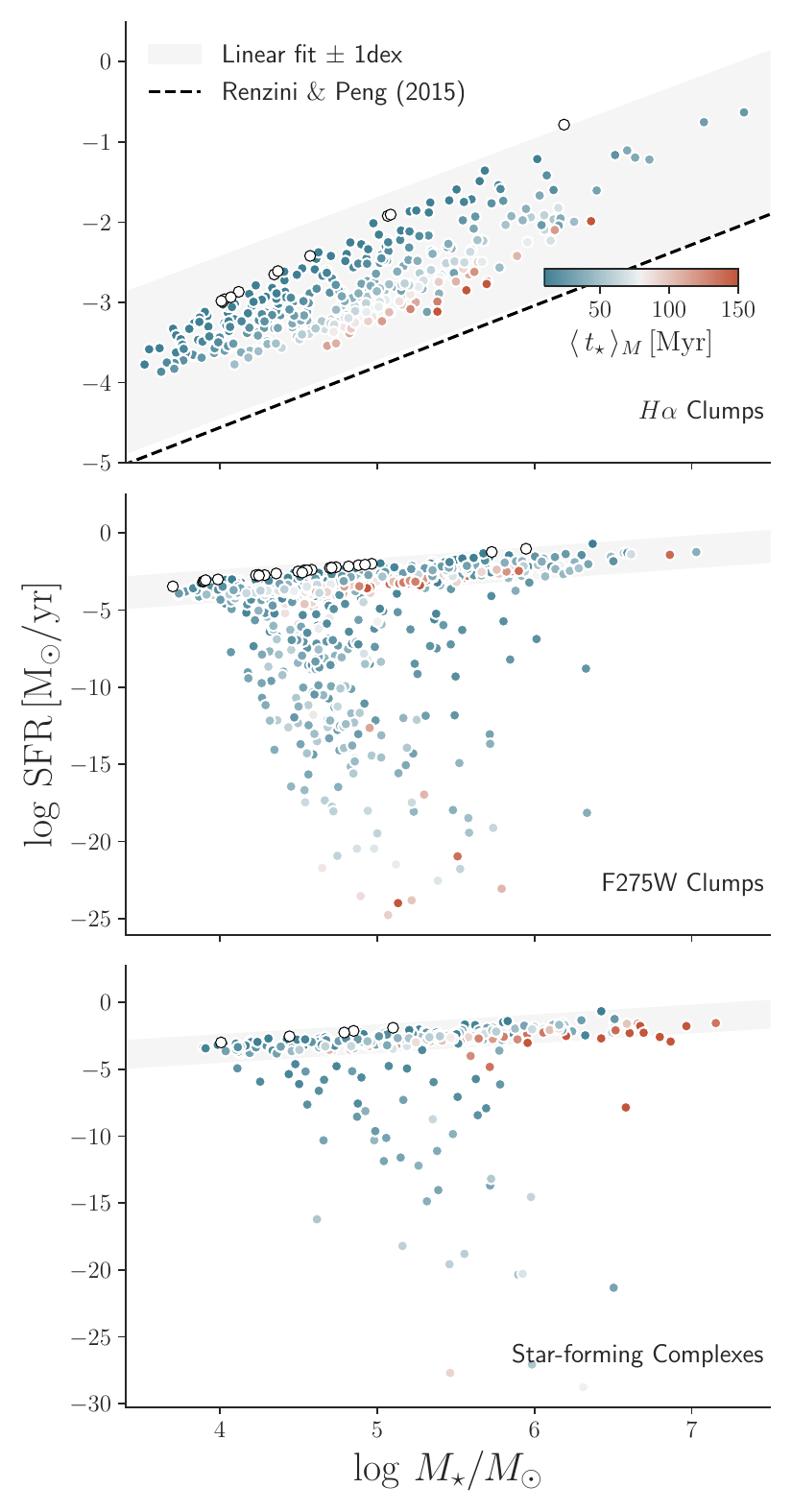}
\caption{Stellar mass -- $\mathrm{SFR}$ relation for clumps and complexes in our sample. Panels from top to bottom show the relations for $H\alpha$ clumps, F275W clumps and star-forming complexes, respectively. Points are colored according to $\langle\,t_\star\,\rangle_M$, truncating at 10 Myr and 150 Myr, clumps containing only stars younger than 10 Myr form a relation by definition and are shown as white circles. The gray band indicates a $\pm1$ dex range around a linear fit to the $H\alpha$ relation, the dashed line shows a fit to the relation for star-forming galaxies in the Sloan Digital Sky Survey from \cite{Renzini2015}. For F275W clumps and star-forming complexes we set the lower limit of the $y$-axis to the 1\% percentile of the $\mathrm{SFR}$ distribution.}
\label{fig:mass-sfr}
\end{figure}

A widely studied characteristic of the population of star-forming galaxies is the star-forming main sequence in the stellar mass - $\mathrm{SFR}$ plane. This relation is also studied for star-forming clumps, which are shown to follow similar (but not equal) trends \citep{Mehta2021}.
We show the $\log\,M_\star$-$\log\,\mathrm{SFR}$ relation for our sample of clumps and complexes in Fig. \ref{fig:mass-sfr}.

$H\alpha$ clumps (top panel in Fig. \ref{fig:mass-sfr}) form a well defined sequence in the stellar mass - $\mathrm{SFR}$ plane, akin to the one observed for normal star-forming galaxies, with all points lying within $\pm$ 1 dex from a linear fit to the relation ($\log\,\mathrm{SFR/M_\odot\,yr^{-1}} = 0.73 \times \log\,M_\star/M_\odot - 6.4$). White points with black edges in the figure represent clumps containing only stars younger than 10 Myr and are not considered in the linear fit as they form a perfect correlation by definition, marking the maximum possible specific $\mathrm{SFR}$. 
At fixed $\mathrm{SFR}$, age increases towards more massive clumps. This can be partially attributed to the SFH parametrization used in this work (see section \ref{sec:synthesis}), 
but we note that \cite{Zanella2019} has also found (at $z\sim2$) that more massive clumps tend to be older, and
a comparable behavior was observed by \cite{Zanella2015} for simulated clumps in the Schmidt-Kennicutt plane \citep{Kennicutt1989}.

Differences in methodology limit our ability to make strict comparisons with the literature; however, comparisons can be made to some extent as long as their limitations are kept in mind.
It is interesting that the slope of the relation obtained by \cite{Renzini2015} for the global (galaxy-wide) relation of star-forming galaxies in the Sloan Digital Sky Survey \citep[SDSS,][]{SDSS} is remarkably similar (0.76$\pm0.01$) to the one of our linear fit (0.73). 
Although the slopes are similar, the relation from \cite{Renzini2015} lies close to -1 dex from our relation in the $\log\,M_\star$ range of interest.
It is indeed expected for star-forming clumps to be above galaxies in this relation \citep[e.g][]{Mehta2021}, as those represent isolated actively star-forming regions, while for whole galaxies these regions are averaged out with non star-forming regions, decreasing the 
specific SFR.
We note that the difference in SFR indicators between this work and \cite{Renzini2015} (SED fitting and $H\alpha$ luminosity) should not be affecting the comparisons, as the SFRs obtained from {\sc bagpipes} with a 10 Myr timescale are very similar (median absolute difference $\lessapprox0.1$ dex) to those obtained by applying the \cite{Kennicutt1998} conversion from $H\alpha$ luminosity ($L_{H\alpha}$) to $\mathrm{SFR}$ ($\mathrm{SFR}/\mathrm{M_\odot\, y^{-1}}=10^{-41.28}\,L_{H\alpha}/\mathrm{erg\,s^{-1}}$ for a Kroupa IMF as in \citealt{Brinchmann2004}) to either the $H\alpha$ luminosity of the {\sc bagpipes} model spectra or the continuum subtracted photometric $H\alpha$ luminosity from \cite{Eric2023}, corrected by dust using a Calzetti attenuation law and the $A_V$ value for young stellar populations obtained from the {\sc bagpipes} fit. 

For F275W clumps and star-forming complexes (middle and bottom panels in Fig. \ref{fig:mass-sfr}) the behavior in the $\log\,M_\star$-$\log\,\mathrm{SFR}$ plane is more complex, with some clumps/complexes lying around the same correlation found for $H\alpha$ clumps, and other straying away towards lower $\mathrm{SFRs}$.   
29.6\% of F275W clumps and 20.8\% of star-forming complexes are below 1 dex from the relation found for $H\alpha$ clumps. For these objects, mass-weighted ages increase with the distance from the relation, indicating passive aging of stellar population reminiscent of the process of star-formation quenching observed in galaxies.
All objects falling away from the relation are classified as early decliners according to their SFHs (see sec. \ref{sec:t0_tau}), but the youngest early decliners are still within $\pm1$ dex of the relation.
On the other hand, clumps and complexes along the sequence have mass-weighted ages that increase towards higher stellar masses and $\mathrm{SFRs}$ similarly to what is observed for $H\alpha$ clumps.

This same relation was studied by \cite{Vulcani2020b} for the entire sample of stripped galaxies in GASP (which includes the 6 galaxies in our sample), where clumps are detected from the $H\alpha$ emission in the MUSE data cubes. The authors find a much steeper slope of 1.63. We acknowledge that the results presented here are in tension with the ones of \cite{Vulcani2020b}, which can most likely be ascribed to differences in spatial/spectral resolution, wavelength range and methodology. Understanding how these methodological differences and selection effects work on shaping these difference is an open issue that we defer to future work.


\subsection{Trends with galactocentric distance}\label{sec:distance_trends}

\begin{figure*}[ht]
\includegraphics[width=\textwidth]{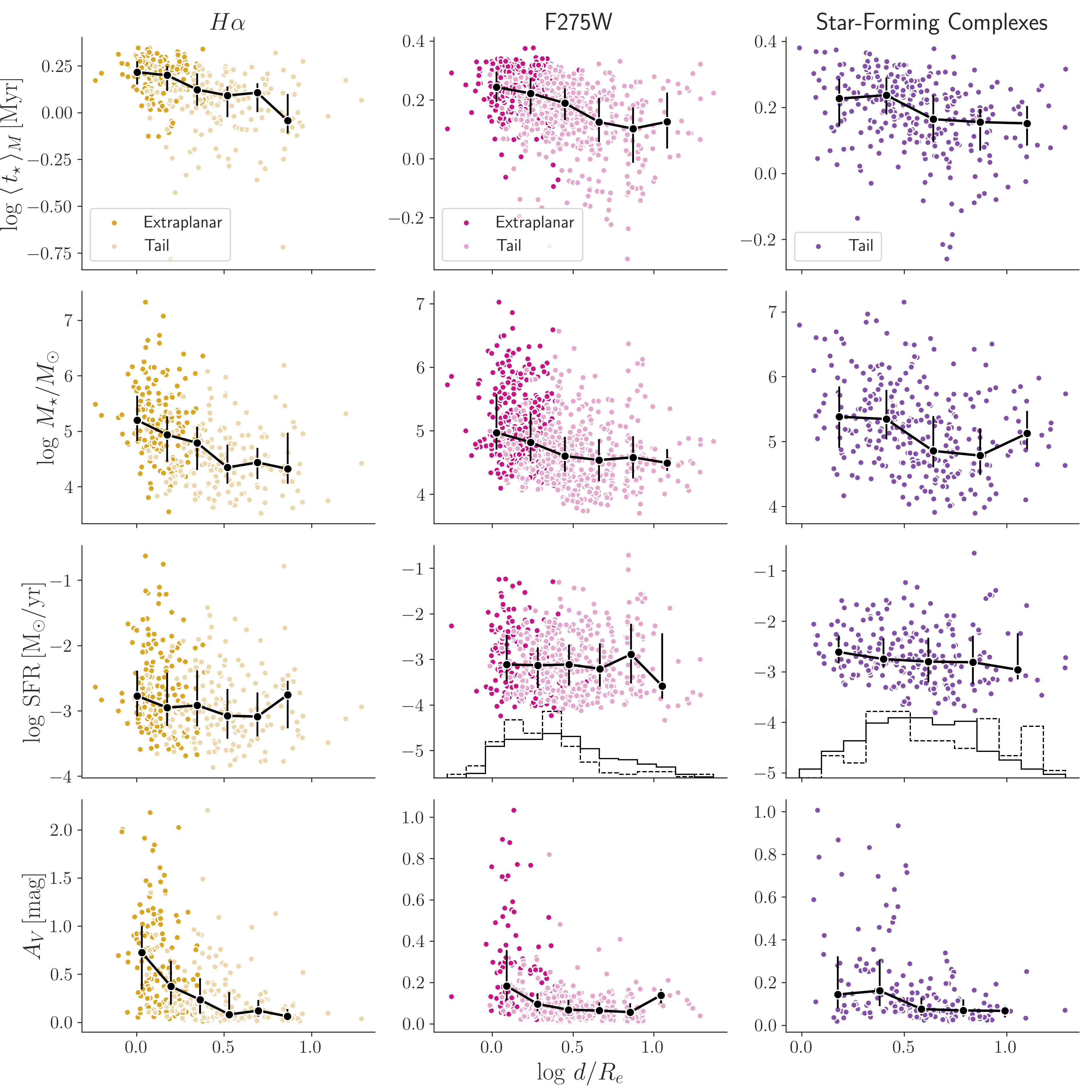}
\caption{$\langle\,t_\star\,\rangle_M$ (top), $\log \, M$ (second row), $\mathrm{SFR}$ (third row) and $A_V$ (bottom) 
of clumps and complexes plotted against the normalized log distance $\log\,d/R_e$ to the center of the host galaxy. The different panels show clumps selected from $H\alpha$ (left) and F275W (middle), as well as star-forming complexes (right).
Black points trace the median values of physical properties in bins of $\log\,d/R_e$ and vertical lines indicate interquartile regions, points are plotted in the center of each bin.
In the third row of panels, only objects within $\pm1$dex of the sequence shown in Fig. \ref{fig:mass-sfr} are included as points, histograms show the distribution of objects below (dashed line) and within (solid line) $\pm1$dex of the sequence. In the bottom row we show only objects that formed a significant fraction of their stellar mass in the past 20 Myr (classified as late decliners in sec. \ref{sec:t0_tau}).}
\label{fig:distance_trends}
\end{figure*}

The properties of star-forming regions at different distances from the stripped galaxy can give us hints about the process of star-formation itself, and also about how stripping occurs in a galaxy.
In Fig. \ref{fig:distance_trends} we plot $\langle\,t_\star\,\rangle_M$, $\log \, M$, SFR and $A_V$ of clumps and complexes in our sample against the log projected distance to the center of the host galaxy, normalized by its effective radius ($\log\,d/R_e$).
We use effective radii measured by \cite{Franchetto2020} using MUSE data.
Not normalizing the radius or using the distance to the edge of the galaxy disk yields very similar (and qualitatively equivalent) trends.

We observe a clear \footnote{Person $R\sim-0.4$ for clumps and $\sim-0.3$ for complexes..} anti correlation of $\langle\,t_\star\,\rangle_M$ with distance for all sub-classes of objects, which is stronger in $H\alpha$ clumps, and slightly weaker for F275W clumps and star-forming complexes; thus, mass-weighted ages get on average younger with increasing distance from the disk.
This is in line with previous works that find similar trends \citep{Fumagalli2011, Kenney2014, Callum2019}. 
We interpret this anti-correlation as due to the effect of ram-pressure on different gas phases.
When the multiphase gas is stripped, atomic gas is easily moved out to large distances from the disk.
On the other hand, the molecular gas, closer to the conditions necessary for star-formation, is less efficiently stripped \citep[see][]{Bacchini2023}.
However, RPS tails are known to host a large amount of diffuse molecular gas \citep{Jachym2019, Moretti2020}.
The age trend that we find suggests that the denser component of the molecular gas is not able to travel long distances before it is turned into stars, which are not susceptible to ram-pressure.
 This locks these early star-forming regions to the vicinity of the disk, while atomic gas continues to be stripped away until it forms new molecular clouds (as the ones observed by \citealt{Moretti2020}) and eventually reaches the conditions for star-formation at large distances from the disk. This interpretation is in line with the results of simulations \cite[see][]{Tonnesen2012}.

We also observe a anti correlation\footnote{Person $R\sim-0.4$ for $H\alpha$ clumps and $\sim-0.3$ for F275W clumps and star-forming complexes.} of stellar mass and distance from the host galaxy, which we interpret as a consequence of the trend with $\langle\,t_\star\,\rangle_M$. Under the assumption that star-formation rate does not drop significantly over the SFH, the older objects have been forming stars for a longer time, and thus have assembled more mass. This assumption is valid for $H\alpha$ clumps as they are mostly late decliners (see sec. \ref{sec:t0_tau}), but less so for F275W clumps and star-forming complexes, for which the SFHs are more diverse.
Moreover, under the interpretation that star-forming regions near the disk originated from the dense molecular gas component of the ISM, 
these regions would also have a larger supply of gas to turn into stars.
Cloud crushing simulations also show that as gas is being stripped and cooled it might also be fragmenting into lower mass clumps, which would also lead to lower stellar masses at larger distances \citep[e.g][]{Abruzzo2023}.

In the third row of panels in Fig. \ref{fig:distance_trends}, 
we plot the $\mathrm{SFR}$ vs. $\log\,d/R_e$
for objects within $\pm1$dex of the  fit to the stellar mass -- SFR relation of $H\alpha$ clumps (see Fig. \ref{fig:mass-sfr}). 
The $\mathrm{SFR}$ values of these objects do not vary significantly with $d/R_e$. 
The $\log\,d/R_e$ distribution
for objects below (by at least 1 dex) a linear fit to $H\alpha$ sequence is traced by the dashed histograms in the same panels, and we also include the $\log\,d/R_e$ distribution of objects within the sequence for comparison.
We also find that there is no striking difference between the $\log\,d/R_e$ distributions of objects that are within or below the sequence defined by $H\alpha$ clumps; indicating that the quenched fraction of F275W clumps and star-forming complexes does not vary with galactocentric distance.

Large values of $A_V$ are observed only in clumps close to the stellar disk, as already seen from the difference between the $A_V$ distributions of tail and extraplanar clumps in Fig. \ref{fig:all_hist}. This supports the idea that dust is not efficiently stripped to large distances from the disk, and that the $A_V$ measured in the {\sc bagpipes} fit might be due to dust that is formed in-situ, or diffuse dust surrounding the galaxy disk.
Another interesting possibility is that dust might be destroyed via sputtering due to the interaction with the ICM \citep{Draine1979, Gutierrez2017, Vogelsberger2019}.

The number of clumps per galaxy is not sufficient to allow meaningful interpretations of all trends (or lack thereof) in a galaxy-by-galaxy basis. Nevertheless, we include a version of Fig. \ref{fig:distance_trends} showing data of individual galaxies in Appendix \ref{sec:distance_galbygal} for reference.

\subsection{Onset of star-formation}

Another interesting parameter when studying jellyfish galaxies is the age of the oldest stars in the tails. Under the assumption that star-formation starts immediately after stripping and stars remain detectable for long enough, the age of the oldest stars in the tails would correspond to the stripping timescale. Since star-formation can take some time to occur and stellar associations fade, the age of the onset of star-formation in the clumps
 serves as a lower (younger) limit for the onset of stripping.
As introduced in sec. \ref{sec:t0_tau}, the age of the oldest stars in our SFH models corresponds to the value of $t_0^\star$ (related to $t_0$ in equation \ref{eq:sfh}). We note that $t_0^\star$ can be degenerate with $\tau$ and is typically not as well constrained as $\langle\,t_\star\,\rangle_M$.

\begin{figure}[ht]
\includegraphics[width=\columnwidth]{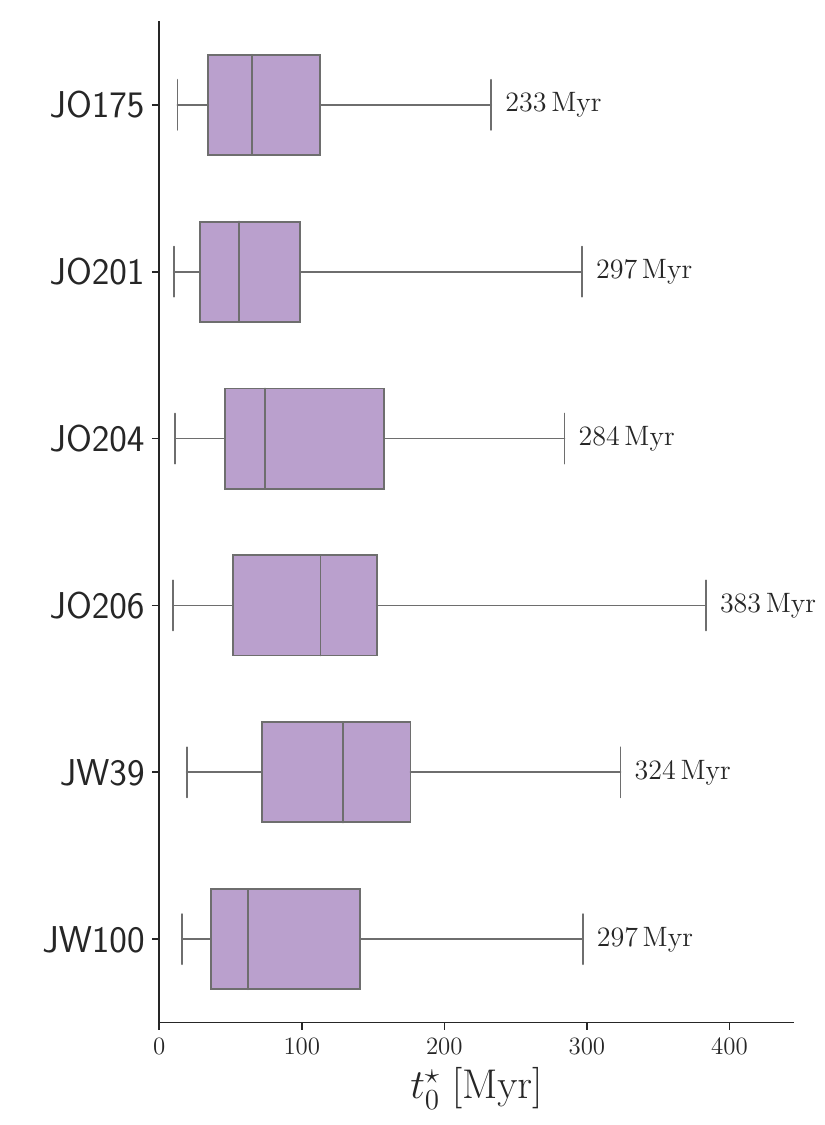}
\caption{Box plots tracing the distribution of age of the oldest stars in star-forming complexes. Boxes span the interquartile regions, horizontal lines in the boxes indicate the median and whiskers mark the 1\% and 99\% percentiles of the distributions.}
\label{fig:onset}
\end{figure}

A similar analysis has been performed for JO201 by \cite{Callum2019} using MUSE data from the GASP project and the \textsc{sinopsis} spectral synthesis code \citep{Fritz2017}. The authors find that the the star-forming regions detected in $H\alpha$ with MUSE in the tail that galaxy are typically younger than 800 Myr (with two outliers as old as 2 Gyr).

We probe the timescale of the onset of star-formation using the $t_0^\star$ value of the star forming complexes. These regions are the oldest in our sample, and also the largest ones which makes them more comparable to the MUSE regions studied by \cite{Callum2019} at a lower spatial resolution. 
In Fig. \ref{fig:onset}, we show box plots tracing the distributions of $t_0^\star$ for each galaxy in our sample, with whiskers indicating the 1\% and 99\% percentiles.
The 99\% percentiles of $t_0^\star$ are very similar between galaxies, varying from 284 Myr for JO204 to 383 Myr for JO206. These values are comparable to that of \cite{Fumagalli2011}, who constrain the ages of blobs in the tail of IC 3418 to be younger than 400 Myr.

Taken at face value, the similarity of the maximum $t_0^\star$ in different galaxies points to a scenario where the stripping timescales do not strongly depend on galaxy properties and surrounding environment.
However, it likely that the $t_0^\star$ distributions are limited by our ability to detect the oldest complexes, as very old stellar associations that quenched star-formation long ($\sim300$ Myr) ago may not be detectable in either F275W or $H\alpha$ and thus would not be included in our sample.
Moreover, galaxies in our sample were selected because all of them show strong stripping features, and the the similar timescales might be a consequence of this selection.

\section{Structure and evolution of star-forming complexes}\label{sec:complexes}

The star-forming clumps studied in this work are rarely found in isolation, and are usually part of larger structures (star-forming complexes). In this section, we focus on the morphology and evolution of star-forming complexes as a whole. We stress that the present section focuses only on complexes that are considered resolved (see section \label{sec:detection} for details.

\subsection{No conspicuous evidence for morphological evolution}

Star-forming regions under ram-pressure are often observed to be organized in structures called fireballs, where stellar populations of different ages are linearly displaced from one another \citep{Kenney2014, Giunchi2023b}).
Fireballs take shape as ram pressure acts in regions that are continuously forming new stars, shaping their morphology into elongated structures. 
Thus, we expect some degree of morphological evolution for clumps and complexes in our sample.
These morphological changes can be traced by the axial ratio (the ratio $b/a$ between the lengths of the minor and the major axes) of the best-fitting ellipse of each resolved clump or complex. A perfectly round object will have $b/a=1$, and values get closer to zero for more elongated objects.
We note that \cite{Giunchi2023b} has already shown that many of the star-forming complexes in our sample of galaxies are organized in fireball-like configurations. However, it is interesting to revisit this analysis using the information obtained in this work and verify if clumps and complexes undergo any morphological changes as they age and assemble stellar mass.

\begin{figure}[ht]
\includegraphics[width=\columnwidth]{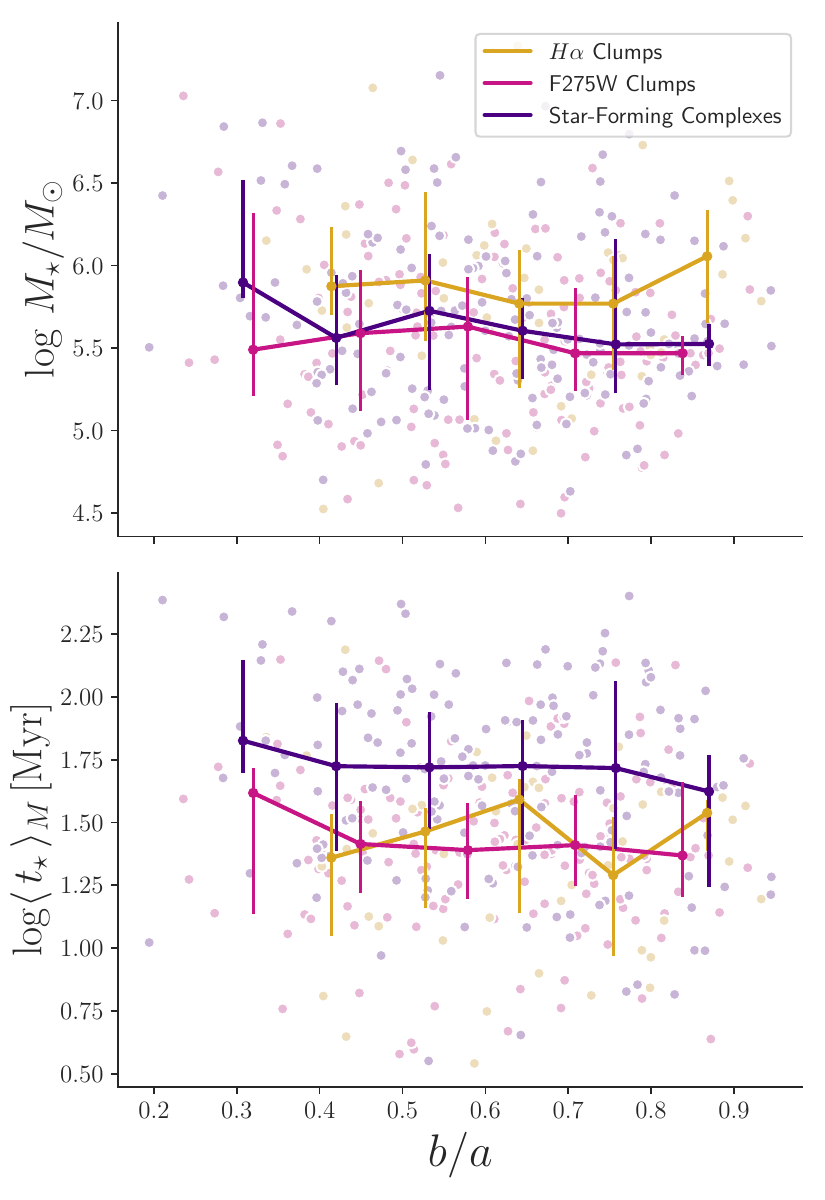}
\caption{Axial ratio ($b/a$) against stellar mass (top) and $\langle\,t_\star\,\rangle_M$ (bottom) for $H\alpha$ clumps (gold), F275W clumps (pink) and star-forming complexes (purple). Median lines are color-coded as the points and error bars represent interquartile regions for each $b/a$ bin.}
\label{fig:axratio_clumps}
\end{figure}

However, when probing this idea (Fig. \ref{fig:axratio_clumps}),
we do not find any significant trend of $b/a$ with stellar mass and $\langle\,t_\star\,\rangle_M$ for clumps or complexes.
We note that the correlations might be stronger than we are able to assess due to the resolution of the observations. Small complexes, that are unresolved in our observations, are less massive and also more likely to be rounder. Higher spatial resolution could potentially populate the lower-right portions of Fig. \ref{fig:axratio_clumps} and possibly reveal stronger trends, although we are unable to extrapolate to which extent this would happen. 
Also, projection effects strongly bias this analysis, as objects that are elongated along the line of sight are seen as round.
Furthermore, star-forming clumps come in a variety of shapes even in the absence of ram-pressure \citep[see simulations from][]{Mandelker2017} so it is possible that these trends are intrinsically weak.

\subsection{A trace of fireballs}

\begin{figure*}[ht]
\includegraphics[width=\textwidth]{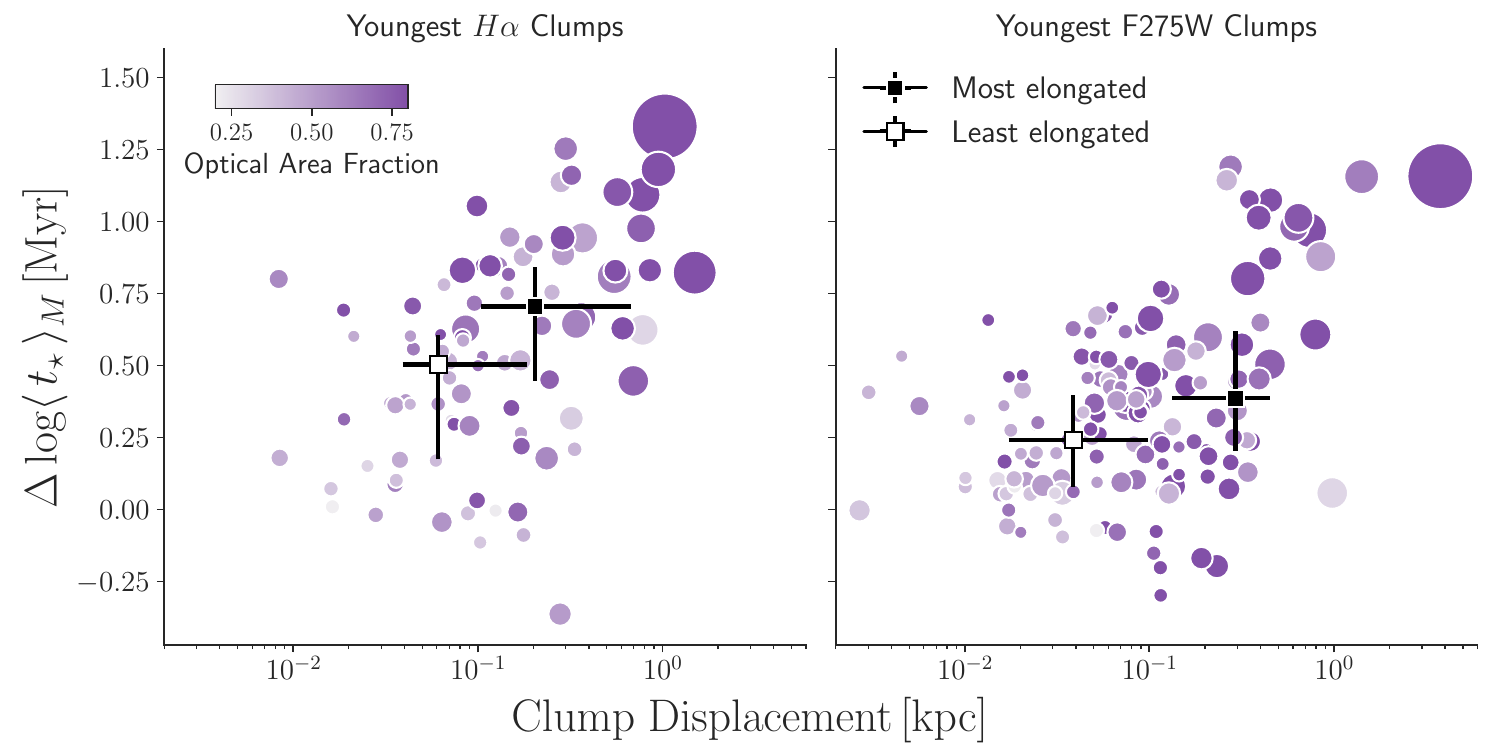}
\caption{Clump displacement as a function of the difference between the log $\langle\,t_\star\,\rangle_M$ of a star-forming complex and the one of the youngest $H\alpha$ (left) and F275W (right) clump embedded in it. 
Color indicates the fraction of the star-forming complex that is covered by optical-only emission, and the size of the points is proportional to the radius of the complex.
Squares indicate the median position of the 20\% most elongated (black) and the 20\% least elongated (white) complexes, with error bars showing the interquartile regions of the distributions.}
\label{fig:age-displacement}
\end{figure*}

To look further into the structure of star-forming complexes, we investigate how the displacement of clumps within a complex correlates with the difference in mass weighted age between clumps and their host complexes ($\Delta \log \langle\,t_\star\,\rangle_M$).
To obtain the measurement of clump displacement, we isolate the region of the complex that is detected only in the F606W filter, i.e the region not covered by any clumps, and calculate the physical distance between the center of this region and the center of each embedded clump. For this analysis, we consider only the youngest clump inside each resolved complex, as these are the ones expected to be more displaced.
The results of this experiment are shown in Fig. \ref{fig:age-displacement}. To aid interpretation we have made the size of the points proportional to the radius of the complexes and colored them according to the area of the complex that is devoid of clumps. We note that several complexes lack either $H\alpha$ or F275W clumps, which is why the number of points in each panel is different.

We observe reasonable linear correlations between clump displacement and $\Delta \log \langle\,t_\star\,\rangle_M$ for both $H\alpha$ (Pearson $R=0.44$) and F275W (Pearson $R=0.41$) clumps. The clumps that contain younger stars (when compared to the age of the complex) tend to be further away from the center of the complex, as expected in a fireball-like structure.
Furthermore, displacement (as well as age difference) is larger in the largest complexes (larger points in the Figure), which also typically have a smaller fraction of their area covered with clumps.

Changes in clump elongation are traced by the black and white squares in Fig. \ref{fig:age-displacement}, which represent the median position of the 20\% most elongated (black) and least elongated (white) complexes, with error bars spanning from the 25th to the 75th percentile of the distributions of $\Delta \log \langle\,t_\star\,\rangle_M$ and displacement.
Although the variation in $b/a$ along the plane of Fig. \ref{fig:age-displacement} is small, 
the difference between these two extreme populations shows that the most displaced clumps are typically in the most elongated complexes.  

\begin{figure}[ht]
\includegraphics[width=\columnwidth]{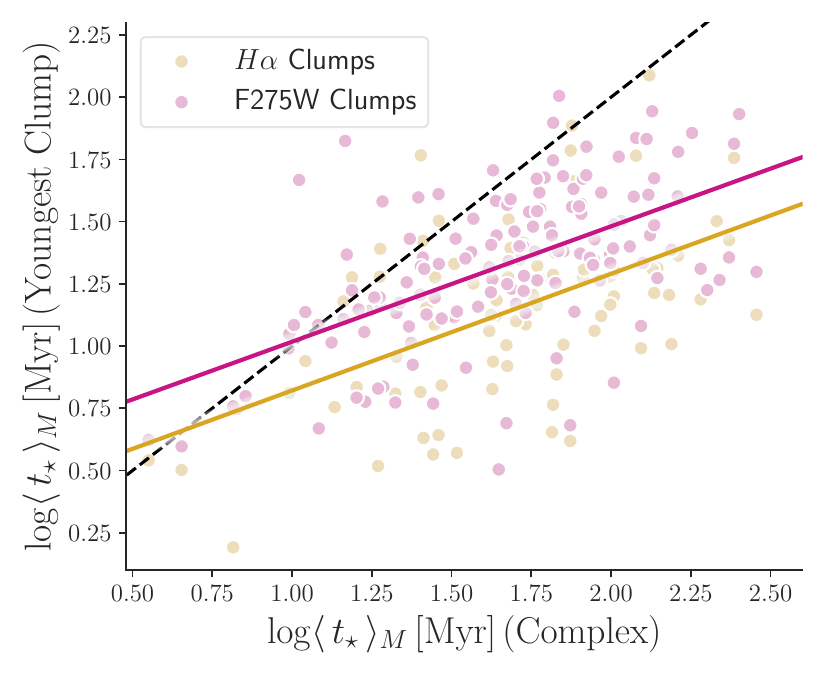}
\caption{$\log \langle\,t_\star\,\rangle_M$ of star-forming complexes against $\log \langle\,t_\star\,\rangle_M$ of the youngest $H\alpha$ (golden) and F275W (pink) clumps inside them. An y=x function is represented by a dashed black line. Solid lines color-coded as the points are linear fits to the relation for the two sub-samples of clumps.}
\label{fig:complex-clump-age}
\end{figure}

We note that the increase in $\Delta \log \langle\,t_\star\,\rangle_M$ shown in Fig. \ref{fig:age-displacement} occurs despite the fact that the clumps in older complexes are also older themselves. We illustrate this in Fig. \ref{fig:complex-clump-age}, where the age of the clumps is plotted against the age of their respective complex for the same objects as in Fig. \ref{fig:age-displacement}. As complexes get older, so do their embedded clumps, but as a sub-linear function of the $\log \langle\,t_\star\,\rangle_M$ of the complex, leading $\Delta \log \langle\,t_\star\,\rangle_M$ to increase as clumps deviate more and more from the identity function (dashed black line in the Figure). 
Moreover, it is noticeable 
that some clumps are above the dashed line in the Figure, being older than their host complex (which can also be seen in Fig. \ref{fig:age-displacement}). For $H\alpha$ clumps this is simply due to uncertainties in the age measurement, while for F275W the difference might also be due to $H\alpha$ clumps in the same complex shifting the overall $\log \langle\,t_\star\,\rangle_M$ to younger ages.

\subsection{Comparisons with similar objects}

Having established the properties of star-forming complexes in the tails of jellyfish galaxies in our sample, we may speculate on possible connections with other objects in the same mass range. 
We note that these comparisons are made under the assumption that star-forming complexes are gravitationally bound structures, which we are not able to confirm with the available data.

\begin{figure}[ht]
\includegraphics[width=\columnwidth]{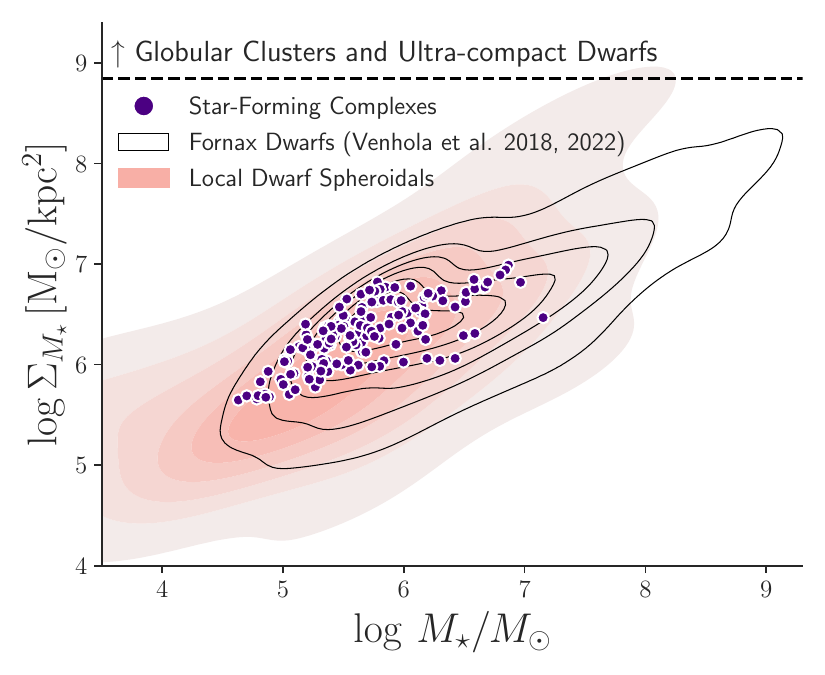}
\caption{$\log\,M_\star$ - $\log\,\Sigma_{M_\star}$ relation for star-forming complexes (purple). Dashed line traces the 5\% percentile of $\log\,\Sigma_{M_\star}$ for globular clusters and ultra-compact dwarfs. Red and black contours trace the distributions of local dwarf spheroidals from \cite{Norris2014} Fornax dwarf galaxies from \citealt{Venhola2018, Venhola2022}.}
\label{fig:sigma_mass}
\end{figure}

In Fig. \ref{fig:sigma_mass} we show the relation between stellar mass and stellar mass surface density for our sample of star-forming complexes, comparing them to objects at similar stellar mass ranges, namely
globular clusters, ultra-compact dwarfs and local dwarf spheroidals from \cite{Norris2014}, as well as dwarf galaxies in the Fornax cluster \citep{Venhola2018,Venhola2022}.
Note that the public catalog from \cite{Norris2014} does not provide a separation between globular Clusters, ultra-compact dwarfs and compact ellipticals, but we exclude the latter by applying a mass cut for $\log M/M_\odot<8$.
The catalog includes additional data from many sources, the relevant references for the objects included in the figure are \cite{Brodie2011, Norris2011, Strader2011, Norris2012, Forbes2013} and \cite{Mieske2013} for globular clusters and ultra-compact dwarfs, and \cite{Walker2009, McConnachie2012} and \cite{Tollerud2012, Tollerud2013} for local dwarf spheroidals. 
For objects in the \cite{Norris2014} sample, we use the stellar masses provided in the catalog, while for the Fornax dwarfs we use the empirical relation between $g-i$ color and mass-to-light ratio derived by \cite{Taylor2011}, neglecting the difference between observed and rest-frame colors (k-correction), which is very small for broad bands at the Fornax redshift.

Globular clusters and ultra-compact dwarf galaxies lie in a mass range quite similar to the one of star-forming complexes. However, these objects have effective radii of up to $\sim$10 pc \citep[e.g][]{Misgeld2011}, while our star-forming complexes have sizes of up to $\sim$1 kpc. 
This difference in size leads to a large difference in stellar-mass surface density.
The dashed line in Fig. \ref{fig:sigma_mass} shows the 5\% percentile of the $\Sigma_{M_\star}$ distribution for globular clusters and ultra-compact dwarfs, which is 2 dex above the 95\% percentile of $\Sigma_{M_\star}$ for our complexes. This difference is much larger than what can be accounted for by methodological differences, and thus star-forming complexes are unlikely to be progenitors of these objects, unless significant morphological evolution takes place.

The position of star-forming complexes in Fig. \ref{fig:sigma_mass} is consistent with the distributions traced by both dwarf galaxy samples being considered. 
However, the local dwarf spheroidals from \cite{Norris2014} reside in widely different environments and are dark matter dominated, while star-forming complexes are expected to remain dark matter free.
On the other hand, the Fornax dwarfs are in an environment consistent with the one of star-forming complexes, and their dark matter content is a topic of ongoing debate \citep{Eftekhari2022, Asencio2023}.

\begin{figure}[ht]
\includegraphics[width=\columnwidth]{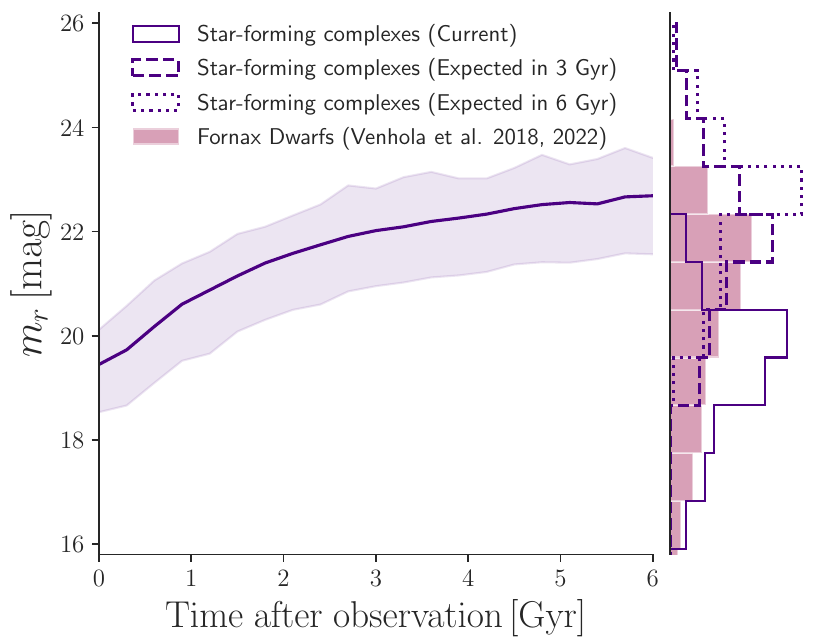}
\caption{Extrapolated r-band apparent magnitude evolution in of star-forming complexes in the next 6 Gyr, placing them at the redshift of the Fornax cluster. The solid line shows the median extrapolated magnitude in each time step and the shaded region traces the 25th and 75th percentiles. Projected histograms show the magnitude distribution for Fornax dwarfs (pink), current magnitude distribution of star-forming complexes (solid purple) and the expected distribution in 3 and 6 Gyr (dashed and dotted purple, respectively). Only complexes older than 50 Myr are included.}
\label{fig:fornax}
\end{figure}

To further compare our star-forming complexes to the Fornax dwarfs, we have placed them at the redshift of the Fornax cluster and extrapolated their SFHs to obtain their expected properties up to 6 Gyr after the observation (note that the magnitudes change very little after 6 Gyr). To ensure relatively stable results, this experiment is performed only for the 121 resolved complexes that are older (in terms of $t_0^\star$) than 50 Myr.
In Fig. \ref{fig:fornax} we show the extrapolated evolution of the r-band apparent magnitude (at the Fornax redshift) of star-forming complexes and compare it with the magnitude distribution of Fornax dwarfs.
The expected magnitudes of star-forming complexes remain consistent with Fornax dwarfs at all simulated time steps, but are mostly situated in the faint end at 6 Gyr. 
This indicates that the remnants of star-forming complexes would stay detectable in nearby clusters such as Fornax for several billion years, possibly contributing to form a population of (dark matter free) cluster dwarf galaxies.

\section{Summary and conclusions}

In this work, we have applied the {\sc bagpipes} SED modeling code to a sample of star-forming clumps and complexes detected by \cite{Eric2023} in the tails and extraplanar regions of 6 GASP jellyfish galaxies observed with HST. 
This allowed us to retrace their star-formation histories and measure their ages, stellar masses, star-formation rates and dust attenuation. 

The typical (median) $\langle \, t_\star \, \rangle_M$ are $\sim$27 Myrs for $H\alpha$ clumps and $\sim$39 Myrs for F275W clumps and star-forming complexes. 
When masking out the clumps inside the complex and leaving only the flux of the region detected in F606W, the median $\langle\,t_\star\,\rangle_M$ is 84 Myrs, and in some cases these objects may include stars older than 300 Myr.
This, combined with the evidence for a diversity of star-formation histories in the sample,
indicates that star-formation can sustained for a long period of time in the tails of jellyfish galaxies.

As expected, $H\alpha$ clumps also have higher star-formation rates when compared to F275W clumps, and their typical SFR is very similar to the one of star-forming complexes if the low SFR tail of the distribution for complexes is excluded.
$H\alpha$ clumps form a well defined sequence in the stellar mass - SFR plane, with a similar slope to (but systematically above) the main sequence of star-forming galaxies. This is not always true for F275W clumps and star-forming complexes. We find that 66\% of F275W clumps and 79\% of star-forming complexes are within $\pm1$ dex from a linear fit to the relation of $H\alpha$ clumps, while the remainder appear to have recently ceased their star-formation activity and are found below 1 dex from the relation.

The stellar masses of objects in our sample vary from $\sim10^{3.5}$ to $\sim10^{7.1}$ $M_\odot$, with median values of 10$^{4.8}$ for $H\alpha$ clumps, 10$^{4.7}$ for F275W clumps and, 10$^{7.1}$ for star-forming complexes. 
Although the stellar masses of star-forming complexes are consistent with the ones of globular clusters, their stellar mass surface densities are lower by 2 dex, which makes it unlikely that the objects studied in this paper may evolve into a population of globular clusters.
On the other hand, their stellar masses and surface densities are compatible with that of dwarf galaxies.
By extrapolating the star-formation histories obtained with {\sc bagpipes} for star-forming complexes, we obtain a $r$-band magnitude distribution that matches the one of dwarf galaxies in the Fornax cluster and show that these objects may remain detectable in nearby clusters for several billion years. Despite the tentative nature of this approach, it serves as indication that the objects formed in the tails of jellyfish galaxies may evolve into a population of dark matter free dwarf galaxies in clusters.

We observe trends of $\langle \, t_\star \, \rangle_M$ and stellar mass with the distance from the center of the host galaxy. Clumps and complexes further away from the galaxy center have younger stellar populations and are less massive.
This can be ascribed to the effect of ram-pressure in different ISM phases, and indicates that the denser component of the molecular gas in the ISM is not able to travel out to long distances from the stellar disk before it is turned into stars, which tend to stay near the stellar disk as they are not susceptible to ram-pressure.
In the meantime, atomic gas continues to be stripped away until it forms new molecular clouds and ultimately lead to the young, low-mass star-forming regions found far away from the galactic disks.
Clumps further away from the galaxy disk are also less attenuated by dust, 
and the distributions of $A_V$ are all concentrated at low ($<0.5\,\mathrm{mag}$) values, except for the population of $H\alpha$ clumps in extraplanar regions.
This indicates that dust is not efficiently stripped to large distances from the host galaxies, or is destroyed in the process.

The difference in the mean age of the stellar populations between the complex and its youngest embedded clump scales with the distance between the clump and the center of the optical emission of the complex.
Although the variation in $b/a$ in our sample is small, it is possible to show that the most displaced clumps are hosted by the most elongated complexes. 
This is consistent with a fireball-like morphology, 
where the cumulative effect of ram pressure moves the star-forming region away from the galaxy disk, leaving behind a trail of slightly older stars.
We also show that clump displacement (and age difference) is smaller for complexes that have young stellar populations overall, and that have a large fraction of their area covered by clumps.
These results show how fireball-like morphologies take shape over time, allowing us to interpret the non-fireball complexes (or the ones with less prominent features) as more recently formed objects that are yet to develop their structure.

\section*{Acknowledgments}
This project has received funding from the European Research Council (ERC) under the European Union's Horizon 2020 research and innovation program (grant agreement No. 833824, GASP project). 
B.~V. and M.~G. also acknowledge the Italian PRIN-Miur 2017 (PI A. Cimatti).
J.F. acknowledges financial support from the UNAM- DGAPA-PAPIIT IN110723 grant, México.

\bibliographystyle{aa}
\bibliography{references}

\appendix

\section{Covarainces between derived parameters}\label{sec:corners}

\begin{figure*}[ht]
\centering
\includegraphics[width=\textwidth]{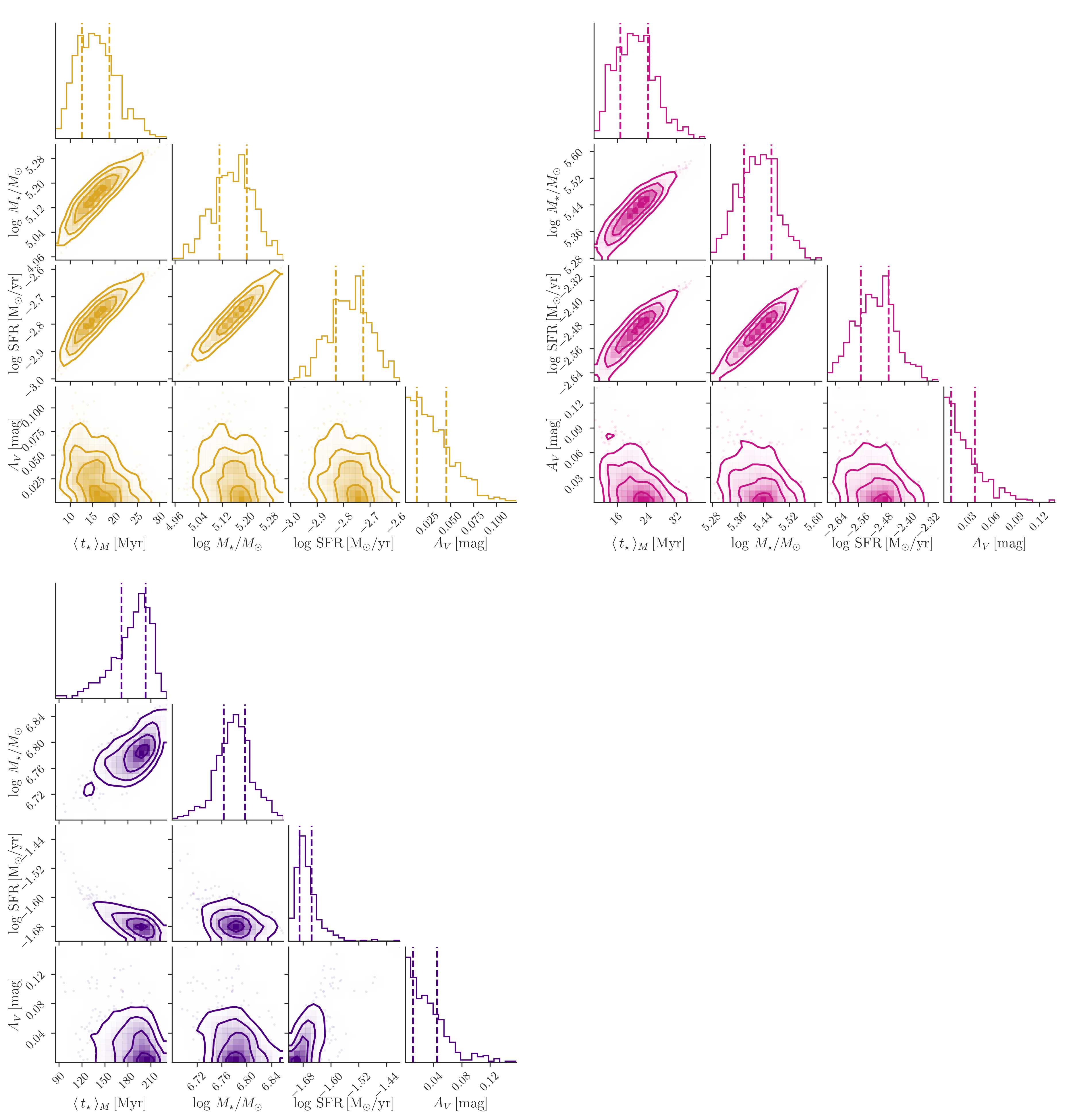} 
\caption{Corner plots showing 2-dimensional projections of the fitted parameter space for the three example fits shown in Fig. \ref{fig:example}. $H\alpha$ clump on the top left, F275W clump on the top right and star-forming complex at the bottom, color-coded as in Fig. \ref{fig:example}. We include the four main variables used in our analysis: $\langle\,t_\star\,\rangle_M$, $\log\,M_\star$, $\log \, \mathrm{SFR}$ and $A_V$.}
\label{fig:corners}
\end{figure*}

Fig. \ref{fig:corners} shows corner plots of the main parameters analyzed in this paper for the three example fits shown in Fig. \ref{fig:example}. As expected, there is covariance between the PDFs of parameters derived from the star-formation history, such as age, stellar mass and star-formation rate. However, this covariance is restricted to very small ranges of values and does not diminish the reliability of the method.

\section{Examples of fits rejected in the quality control}\label{sec:bad_fits}

In section \ref{sec:final_sample} we have presented two quality control criteria that were used to remove 
objects with unsatisfactory fits from our sample. These are: (a) the median of the model flux PDF is outside of the error bars of the observations in two or more filters; or (b) the interquartile range of the model flux PDF is fully outside of the error bars of the observation (i.e more than 50\% of the generated models are outside error bars) in one or more filters.

\begin{figure*}[ht]
\centering
\includegraphics[width=0.9\textwidth]{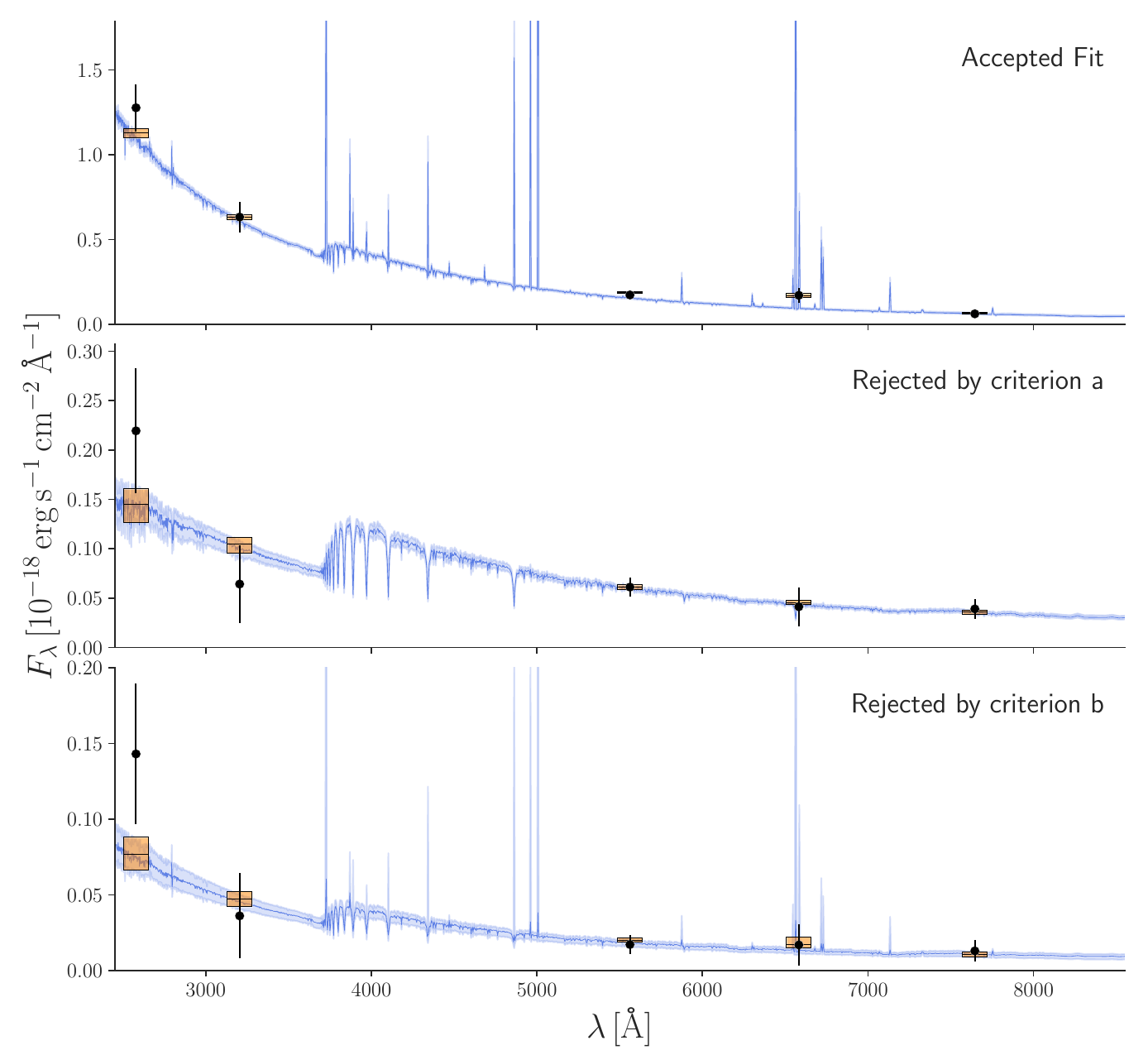}
\caption{Examples of {\sc bagpipes} fits accepted (top) and rejected according to quality control criteria a and b (middle and bottom, see text for details). 
Black points with errorbars show observed photometric fluxes in the 5 HST bands used in this work.
Blue lines show the median of all models generated in the fit, and the shaded regions indicate interquartile ranges.
Rectangles show the interquartile ranges of the modelled photometric fluxes, with horizontal lines indicating their median.}
\label{fig:goodbadugly}
\end{figure*}

In Fig. \ref{fig:goodbadugly}, we provide examples of 
F275W clumps in JO201 
that were rejected according one of the quality control criteria, and also of a fit considered good (although not perfect).
Note that the criteria are somewhat intertwined, 
and work together to establish some degree of tolerance. If one of the median model fluxes is outside error bars but 
the interquartile range still intersects with the error bar and
all others filters are well fitted, the fit is accepted; this case is exemplified in the top panel of Fig. \ref{fig:goodbadugly}.
In the examples, the filters that lead the object to fail the quality control are the two bluest ones, which is the most common case.

\section{Trends (or lack thereof) of properties with galactocentric distance for individual galaxies}\label{sec:distance_galbygal}

\begin{figure*}[ht]
\centering
\includegraphics[width=0.95\textwidth]{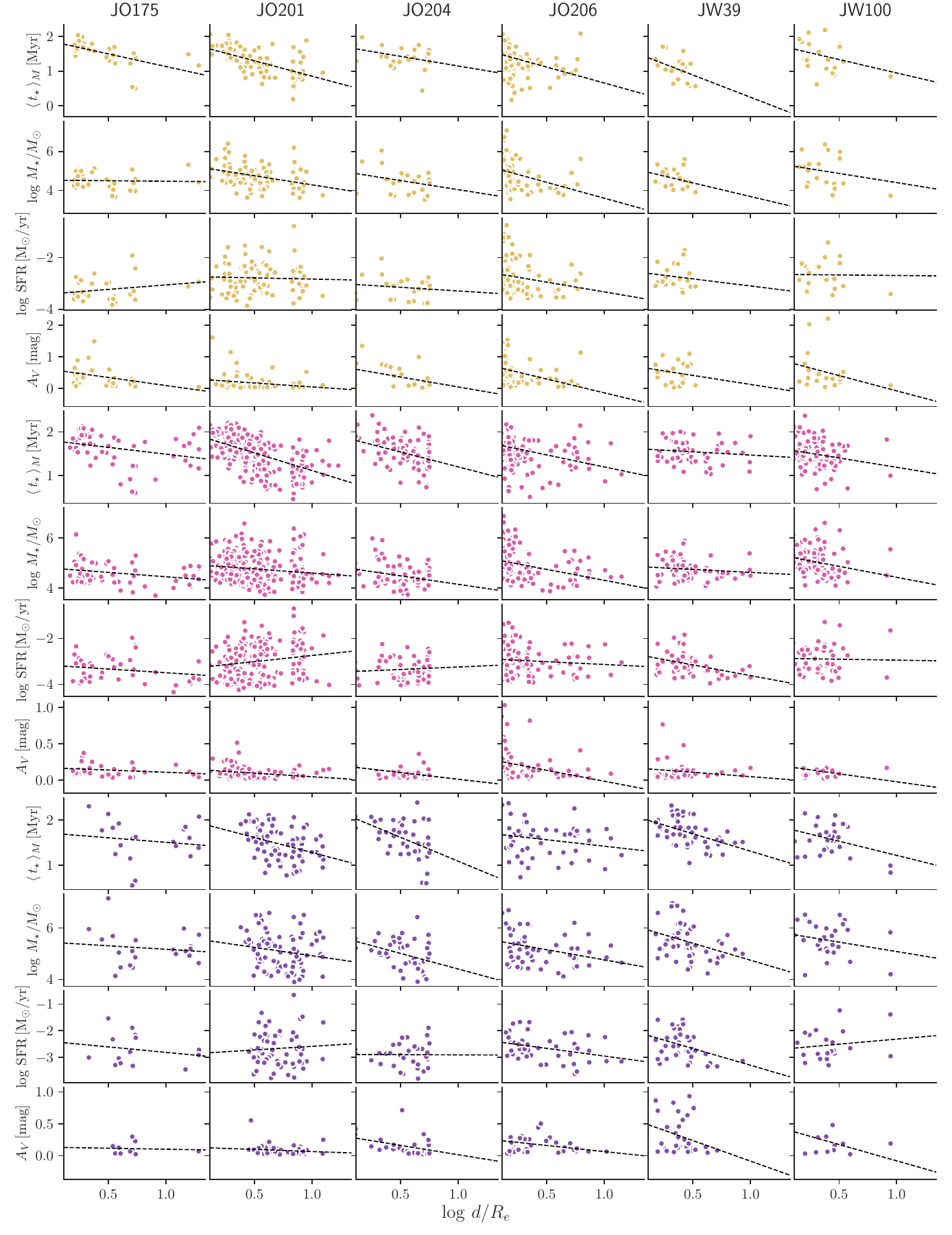}
\caption{$\langle\,t_\star\,\rangle_M$, $\log \, M$, $\mathrm{SFR}$ and $A_V$ 
of clumps and complexes plotted against the normalized log distance $\log\,d/R_e$ to the center of the host galaxy. Each column of panels corresponds to a different galaxy. 
Top panels show data for $H\alpha$ clumps (gold), middle panels show F275W clumps (pink) and bottom panels show star-forming complexes (purple).
Black dashed lines show linear fits to the relations.
For $\mathrm{SFR}$ and $A_V$ we use the same selection as in Fig. \ref{fig:distance_trends}.}
\label{fig:distance_galbygal}
\end{figure*}

In section \ref{sec:distance_trends}, we have explored trends of the properties derived in this work with the distance from the clumps and complexes to the center of the host galaxy. For reference, we show the same trends for individual galaxies in Fig. \ref{fig:distance_galbygal}.

\end{document}